\documentclass[10pt,preprint,a4paper]{aastex}
\usepackage{graphics,epsf}
\usepackage{amsmath}                
\usepackage{amsfonts}               
\usepackage{amssymb}                
\usepackage{epsfig}                 
\usepackage{subfigure}

\def \cm{~\rm{cm}}
\def \s{~\rm{s}}
\def \km{~\rm{km}}

\def \K{~\rm{K}}
\def \g{~\rm{g}}

\def \AU{~\rm{AU}}
\def \erg{~\rm{erg}}

\def \yr{~\rm{yr}}
\def \pc{~\rm{pc}}

\def \days{~\rm{day}}

\title{FORMING EQUATORIAL RINGS AROUND DYING STARS}

\author{Muhammad Akashi\altaffilmark{1}, Efrat Sabach\altaffilmark{1}, Ohad Yogev\altaffilmark{1}, \& Noam Soker\altaffilmark{1}}

\altaffiltext{3}{Department of Physics, Technion -- Israel
Institute of Technology, Haifa 32000 Israel;
akashi@physics.technion.ac.il; efrats@physics.technion.ac.il;
soker@physics.technion.ac.il.}

\begin{document}

\begin{abstract}
We suggest that clumpy-dense outflowing equatorial rings around
evolved giant stars, such as in supernova 1987A and the Necklace
planetary nebula, are formed by bipolar jets that compress gas
toward the equatorial plane. The jets are launched from an
accretion disk around a stellar companion. Using the FLASH
hydrodynamics numerical code we perform 3D numerical simulations,
and show that bipolar jets expanding into a dense spherical shell
can compress gas toward the equatorial plane and lead to the
formation of an expanding equatorial ring. Rayleigh-Taylor
instabilities in the interaction region break the ring to clumps.
Under the assumption that the same ring-formation mechanism
operates in massive stars and in planetary nebulae, we find this
mechanism to be more promising for ring formation than mass loss
through the second Lagrangian point. The jets account also for the
presence of a bipolar nebula accompanying many of the rings.
\end{abstract}

{\bf Key words:} binaries: general � stars: AGB and post-AGB -
stars: jets - stars: mass-loss - (ISM:) planetary nebulae: general

\section{INTRODUCTION}
\label{sec:intro}

Many nebulae around evolved stars posses a dense equatorial ring.
These include both progenitors of core collapse supernovae
(CCSNe), e.g., SN~1987A, and lower mass planetary nebulae (PN) and
pre-PN, e.g., the Necklace PN (PN~G054.2-03.4). There is no
consensus yet on the formation mechanism of the equatorial ring
and its relation to the bipolar nebula found in all these systems.

The inner (equatorial) ring of 1987A is presently energized by the
collision of the SN ejecta, and it can be clearly resolved into
many dense clumps (e.g., \citealt{Larssonetal2013,
Zanardoetal2013}). The radius of the ring is $0.21 \pc$
\citep{Plaitetal1995}, and its total ionized mass is $0.06
M_\odot$ \citep{Mattilaetal2010}. The center of the ring is offset
by $2 \%$ (of its radius) from the SN explosion
\citep{Sugermanetal2005}. The total mass of the circumstellar
matter (CSM) nebula of SN 1987A is $1.7 M_\odot$
\citep{Sugermanetal2005}. This implies that the mass in the
equatorial ring is a small fraction of the nebular mass (even if
the mass in the equatorial ring is several times the ionized mass
in the ring). For a recent summary of the CSM properties of SN
1987A see \cite{Potteretal2014}. \cite{Smithetal2013} and
\cite{Smith2007} find the equatorial ring around the B-type
supergiant SBW1 and that around the luminous blue variable (LBV)
candidate HD 168625, respectively, to be similar in some respects
the inner ring of SN 1987A (see also \citealt{Tayloretal2014}).

\cite{Brandneretal1997} point to some similarity between the ring
nebula of the blue supergiant Sher~25 and the hourglass shape of
the CSM of SN~1987A. For example, the ring in Sher~25 expands at
$30 \km \s^{-1}$, compared with $\sim 10 \km \s^{-1}$ in SN~1987A.
Both the ring and the bipolar lobes of Sher~25 were formed in a
brief and violent major outburst with a very high mass-loss rate
about $6600 \yr$ ago \citep{Brandneretal1997}. In SN~1987A, on the
other hand, based on dynamical timescales, the inner ring is $\sim
10^4 \yr$ younger than the outer rings. The ionized mass in the
hourglass nebula of Sher~25 is $\sim 0.3-0.6 M_\odot$, while that
in the ring is only $0.01-0.1 M_\odot$. \cite{Tayloretal2014} did
not find an interacting or close stellar companion to Sher~25. In
the scenario proposed in the present study a binary companion had
entered the envelope of Sher~25 about $6600 \yr$ ago.

The Necklace PN has been studied by \cite{Corradietal2011} and its
central binary system was further explored by
\cite{Miszalskietal2013}. The ring expands with a velocity of $28
\km \s^{-1}$, and the total ionized mass of the nebula, most of it
in the ring, is $0.06 \pm 0.03 M_\odot$. The ring was ejected
$\sim 5000 \yr$ ago and the two opposite polar caps, which were
most likely formed by two opposite jets, are $\sim 9000-13,000
\yr$ old \citep{Corradietal2011, Tocknelletal2014}. The ring does
not seem to posses perfect symmetry around the central star; there
seems to be more mass to the north-west direction. The orbital
period of the central binary system is 1.16~day, which indicates
that the system went through a common envelope (CE) evolution. The
total mass in the ring, $\sim 0.06 M_\odot$, is much smaller than
any mass ejected in the CE process that shrank the binary orbit. A
somewhat similar structure of bipolar nebulae and a ring is seen
around post-AGB stars \citep{Bujarrabaletal2013a}, e.g., the Red
Rectangle \citep{Cohenetal2004, Bujarrabaletal2013b}. However, in
the Red Rectangle the ring is in a Keplerian rotation
\citep{Juraetal1995}, there is an outflowing disk, and the
secondary star did not enter a CE phase \citep{vanWinckel2003}.
This type of systems will be studied in a forthcoming paper.

We note the following common properties to the equatorial rings in
the Necklace and most of the equatorial rings of the massive stars
with CSM rings, in particular SN 1987A. (1) The general nebular
structure composed of the equatorial ring and a bipolar nebula,
e.g., a nebula with prominent two outer rings in SN~1987A
\citep{Sugermanetal2005}, and polar clumps (formed by jets) in the
Necklace. Other PNe show a clumpy ring in the equatorial plane of
a bipolar structure, e.g., NGC~6309 \citep{Rubioetal2015}. (2) The
equatorial ring is broken to many clumps. (3) Displacement from a
perfect axisymmetry around the star. This suggests a short
interaction time, or interaction in an eccentric orbit. A short
interaction time, say few orbital periods or less, suggest
instability of the orbital motion. We will examine the possibility
of the system being in a Darwin unstable phase during the
formation of the ring. (4) The mass in the ring is a small
fraction of that in the nebula. (5) Large parts of the polar
regions of the nebula, both in SN 1987A and the Necklace, were
formed before the ring. However, in Sher~25 the bipolar nebula and
ring have been ejected together and during a short time
\citep{Brandneretal1997}. As well, it seems that there are bipolar
regions connected to the equatorial ring in SN 1987A
\citep{Sugermanetal2005} and in SBW1 \citep{Smithetal2013}. The
relation to a bipolar structure and the small mass of the ring
suggest that the ring formation comes with the formation of a
bipolar nebula. (6) It is commonly accepted that in all systems,
PN and massive blue giants, a fast wind, $\sim 500-2000 \km
\s^{-1}$, was blown shortly after the ejection of the ring.

There is also one significant difference between the systems. In
the Necklace the CE process had removed the entire envelope, while
in SN~1987A and the other massive stars a massive stellar envelope
had survived. Hence, under our assumption of a common formation
mechanism, the ring in the Necklace was not formed from the final
ejection of the CE.

Let us add here two more systems, the born-again planetary nebulae
A30 and A78 (\citealt{Fangetal2014} for a most recent paper and
references therein). Both PNe show in their inner region an
equatorial ring of knots and faster polar outflows, that
\cite{Fangetal2014} claim to have been ejected in one event in
each PN. The polar knots can reach velocities of $\ga 100 \km
\s^{-1}$. A fast wind ($>1000 \km \s^{-1}$) blown after the
ejection of the ring and polar outflows interacted with the knots
\citep{Fangetal2014}. A late fast wind is common to SN~1987A and
the Necklace as well. Such a wind can clean the inner region from
low density slowly-moving gas, e.g., gas between the dense
equators to condense polar regions. There is no indication yet for
binary systems at the center of these two PNe.

 Another interesting object is the classical nova V959 Mon, for
which \cite{Chomiuketal2014} argue for equatorial outflow and a
faster polar outflow.  A clear bi-lobed morphology appears $\sim
100~$days past the detection of $\gamma$-ray emission.
\cite{Chomiuketal2014} mention that a fast central wind can be
focused to the polar directions by the dense equatorial gas. We
note that it is quite possible that two opposite jets were
launched by the WD few months after outburst, as has been
suggested by \cite{Retter2004} for nova V1494 Aql. In this paper
we follow \cite{SokerRappaport2000} and raise the possibility that
jets compress an equatorial outflow, rather than an equatorial
outflow that focuses a polar outflow, in nova V959 Mon and the
other objects discussed here.

Many models have been proposed for the formation of equatorial
rings, both single-star models, e.g., \cite{Chitaetal2008}, and
binary models, e.g., \cite{MorrisPodsiadlowski2009} who studied a
merger model for the triple ring system of SN~1987A.
 We instead take the view that such nebulae are shaped by jets (e.g.,
\citealt{Soker2007}), and lobes inflated by the jets can compress
an equatorial disk or ring \citep{SokerRappaport2000,
AkashiSoker2008}. We will use the common properties to examine a
common formation mechanism of the equatorial rings in the systems
listed above.

\cite{Chitaetal2008} run a bipolar outflow into a thin shell.
Their study has some common features with the mechanism proposed
by \cite{SokerRappaport2000} and simulated by
\cite{AkashiSoker2008}. \cite{Chitaetal2008} were aiming at
forming several rings outside the equatorial plane. At late times
they have no equatorial ring, but only rings above and below the
plane. Their equatorial ring is observed only at early times, and
it is not formed by compression of gas toward the equatorial plane
by jets. The mechanism studied here is based on jets that compress
an equatorial outflow, as outlined in \cite{SokerRappaport2000},
and should be considered different in many respects from that
studied by \cite{Chitaetal2008}. 

We will differ from \cite{AkashiSoker2008} by considering jets-CSM
interaction much closer to the binary systems in a process that
might also be observed as an Intermediate-Luminosity Optical
Transient (ILOT; also termed: Intermediate-Luminosity Red
Transient; Red Transient; Red Nova; Luminous Red Nova). ILOTs are
eruptive events with a typical decay time of weeks to years and
with a peak luminosity between those of novae and supernovae (e.g.
\citealt{Bond2009, KashiSoker2010b, Kasliwal2011,
Tylendaetal2013}). Several studies in recent years pointed to some
connections between different ILOTS, e.g, between NGC~300OT and
pre-planetary nebulae (PNe) \citep{Prieto2009}, between NGC~300OT
and the Great Eruption of $\eta$ Car, \citep{Kashietal2010}, and
between NGC~300OT and the PN NGC~6302 and the pre-PNs OH231.8+4.2,
M1-92 and IRAS~22036+5306 \citep{SokerKashi2012}.
\cite{SokerKashi2012} proposed that the lobes of some (but not
all) PNs and pre-PNs were formed in a several month-long ILOT
event (or several close sub-events). This process was recently
simulated by \cite{AkashiSoker2013}, who also summarize the
expected properties of the nebular parts of a PN that are formed
by an ILOT event (in general not all nebular components are formed
by an ILOT). \cite{BoumisMeaburn2013}, for example, raised the
possibility that the lobes of the PN KjPn~8 were formed in an ILOT
event.

In section \ref{sec:Darwin} we examine the possibility that the formation episode of the rings,
either by mass loss from the second Lagrangian point ($L_2$) or by jets,
occurred during a relatively short phase of Darwin instability.
In section \ref{sec:L2} we analyze the mass lose through the
second Lagrangian point ($L_2$) and conclude that it is unlikely by itself to explain the equatorial rings.
In section \ref{sec:hydro} we use a 3D hydrodynamical code to simulate ring
formation by a brief jet-launching episode, that might lead also
to an ILOT. Our summary is in section \ref{sec:summary}.

\section{DARWIN UNSTABLE PHASE}
\label{sec:Darwin}

We consider a binary system composed of a primary giant star and a
more compact secondary star, mainly a main sequence (MS) star. The
giant primary star can be a red giant branch (RGB) star, and
asymptotic giant branch (AGB) star, e.g., in the case of PNe, or a
red supergiant (RSG), e.g., in the case of the progenitor of the
SN~1987A nebula. We study processes where the expanding equatorial
rings mentioned in section \ref{sec:intro} were formed in a
relatively short time, i.e., much shorter than the phase of high
mass loss rate. The high mass loss rate of AGB and RSG stars can
last hundreds to thousands of years, while the brief ring
formation episode is assumed to take place within months to tens
of years, $\sim 0.03-30 \yr$. We are not considering here the
formation of rotating (Keplerian) disks.

We assume that the brief ring formation episode takes place while
the system is in an unstable Darwin phase. Before the ring
formation episode the secondary star brings the giant envelope to
synchronization with the orbital motion (co-rotation), and the the
system is Darwin stable, i.e., $I_B > 3 I_1$. Here $I_B$ and $I_1$
are the moments of inertia of the binary system and the primary
star, respectively. Due to further spiraling-in as a result of
loss of angular momentum and/or due to the expansion of the giant
star, at some stage the system becomes Darwin unstable.

We can constrain the mass of the secondary star under these
assumptions when interaction starts. Tidal interaction becomes
strong when the primary radius becomes $R_1 \ga 0.25 a_0$, where
$a_0$ is the orbital separation when tidal interaction starts to
dominate the orbital evolution \citep{Soker1996}. We demand that
by the time the binary system has spiraled-in to $a_i \ga 1.5 R_1
$ the primary envelope has reached co-rotation (the derivation is
not sensitive to the exact value in the range of $a_i \simeq 1.2-2
R_1 \simeq 0.3-0.5 a_0$). The moment of inertia of the giant
envelope is taken to be $I_1 \simeq \eta M_{\rm env} R^2_1$, where
$M_{\rm env}$ is the convective envelope mass and $R_1$ is the
giant radius. For AGB stars $\eta \simeq 0.2-0.24$. For the
progenitor of SN~1987A during its RSG phase we run a model with an
initial mass of $20 M_\odot$ using the Modules for Experiments in
Stellar Astrophysics (MESA), version 6596 \citep{Paxtonetal2011}.
At its RSG phase $20,000 \yr$ before explosion its mass is $16.4
M_\odot$, its radius is $R_1=1130 R_\odot$, the convective
envelope mass is $M_{\rm env} = 9.3 M_\odot$, and $\eta=0.22$.

We will consider cases where $M_2 \ll M_1$. The above condition to
reach synchronization at $a > a_i \simeq 1.5 R_1 \simeq 0.4 a_0$
gives a constraint of
\begin{equation}
M_2 \ga 0.7 \eta M_{\rm env-t} \simeq 0.15 \left(
\frac{\eta}{0.22} \right)M_{\rm env-t} , \label{eq:M2min}
\end{equation}
where $M_{\rm env-t}$ is the giant envelope mass when tidal
synchronization takes place. The Darwin instability might occur
before or after the primary fills its Roche lobe. We demand that
the system enters the Darwin instability when the secondary is
outside the primary, but not too close, e.g., $a_D \simeq 1.5R_1$.
The condition reads
\begin{equation}
M_2 \la 0.3 \left( \frac{\eta}{0.22} \right) \left( \frac{a_D}{1.5
R_1} \right)^{-2} M_{\rm env-D} , \label{eq:M2max}
\end{equation}
where $M_{\rm env-D}$ is the giant envelope mass when Darwin instability occurs.
Over all, the scenario of ring formation during a Darwin unstable phase reads
\begin{equation}
0.15 \left( \frac{\eta}{0.22} \right)  \la \frac{M_2}{M_{\rm env}}
\la 0.3 \left( \frac{\eta}{0.22} \right), \label{eq:M2}
\end{equation}
where we assumed that the envelope mass did not change much
between the tidal synchronization and Darwin instability phases.

For the progenitor of SN~1987A condition (\ref{eq:M2}) reads $1.4
M_\odot \la M_2 \la 2.8 M_\odot$. In the merger model of
\cite{MorrisPodsiadlowski2009} the secondary mass is $5 M_\odot$,
and most of the mass is ejected in mid-latitudes, not in the
equatorial plane. \cite{ChevalierSoker1989} required a secondary
mass of $M_2 \simeq 1 M_\odot$ to account for the deformation of
envelope after the merger. Considering the uncertainties in the
above estimates of the binary companion mass, the mass range found
by us here can accommodate these models. We emphasize that for the
progenitor of SN~1987A we consider a complete rapid merger when
the giant core is made out of CO, unlike the incomplete merger
discussed by \cite{Podsiadlowskietal1990}.

For a low mass AGB star with an envelope mass of $\sim 0.5-1
M_\odot$ near the tip of the AGB, condition (\ref{eq:M2}) reads
$M_2\simeq 0.08-0.3 M_\odot$. This mass range is smaller than the
secondary mass range of $0.4-1 M_\odot$ mentioned by
\cite{Miszalskietal2013} for the Necklace. We later present our
view that the ring in the Necklace was formed in a post-CE event,
like a born-again object.

We now turn to examine two processes that can operate more
efficiently during a Darwin unstable phase and lead to equatorial
mass outflow: the mass loss from the second Lagrangian point $L_2$
(section \ref{sec:L2}), and the launching of two opposite jets
(section \ref{sec:hydro}). The later can work efficiently even if
the system is Darwin stable.

\section{EXAMINING EQUATORIAL MASS LOSS}
\label{sec:L2}

We consider a binary system where the primary of mass $M_1$ losses
mass to a secondary star of mass $M_2$. The orbit is circular with
an orbital separation $a$, and the mass ratio $q \equiv
M_2/M_1<1$. Part or all of this mass is lost through the second
Lagrangian point $L_2$ located behind the secondary star; the
secondary is the accretor in this setting. \cite{Livioetal1979}
already discussed the formation of asymmetrical PNe by mass loss
through $L_2$. We here repeat those calculations and present them
in a way appropriate for our goals. We first consider the angular
momentum of the mass leaving through $L_2$, and then consider the
energy of this mass.

The mass loss process from $L_2$ has been discussed in many
papers, e.g., \cite{Chesneauetal2014} for a recent paper on
V838~Mon, but in many cases for rotating disks rather than
expanding ones (e.g., \citealt{Millouretal2011, Pletsetal1995}).
We here discuss only expanding rings/disks that escape from the
binary system.

\subsection{Angular momentum at $L_2$}
\label{subsec:AML2}

\cite{Livioetal1979} give the value of the specific angular
momentum of mass leaving at $L_2$ as
\begin{equation}
j=\nu \frac{J_B}{M_B},
\label{eq:nu1}
\end{equation}
where $J_B$ and $M_B=M_1+M_2$ are the total angular momentum and
mass of the binary system. Based on the results of \cite{Lin1977},
\cite{Livioetal1979} take $\nu \simeq 1.7 ({M_B}/{\mu})$, where
$\mu \equiv M_1 M_2/(M_1+M_2)$ is the reduced mass of the binary
system,. For mass at $L_2$ the value is
\begin{equation}
\nu = \left(\frac{x_{\rm cm-L}}{a} \right)^2 \frac {M_B}{\mu} <
1.7  \frac {M_B}{\mu}, \label{eq:nu2}
\end{equation}
where  $x_{\rm cm-L}$ is the distance of the center of mass to
$L_2$, and $a$ is the orbital separation. The reason the value at
$L_2$ is somewhat smaller than the final value of the angular
momentum, by $\sim 10 \%$ for the parameters relevant to us, is
that the binary system exerts toque on the mass leaving $L_2$
\citep{Lin1977}.

~From angular momentum considerations we can get an expression for
the decrease in orbital separation when ($i$) mass loss occurs
only from $L_2$, ($ii$) we neglect the moment of inertia of the
primary star, and ($iii$) no mass accretion by the secondary star
takes place. Taking a small mass loss such that $q$ does not
change much, the equation reads
\begin{equation}
dJ_B=j ~ d (m_L) =\nu \frac{J_B}{M_B} ~ d (m_L),
 \label{eq:AM1}
\end{equation}
where $m_L$ is the mass lost from $L_2$. Taking $m_L \ll M_1$ such
that only the orbital separation $a$ changes in the expression for
$J_B$, i.e., $dJ_B= (1/2) J_B ~d (\ln a)$, we arrive at the
equation
\begin{equation}
d (\ln a)= 2 \nu \frac{d(m_L)}{M_B} = 3.4 ~\frac{d(m_L)}{\mu},
 \label{eq:AM2}
\end{equation}
where in the second equality we have substitute for $\nu$ from
equation (\ref{eq:nu1}). For $(a_i-a)/a \ll 1$ the solution is
\begin{equation}
\frac{a}{a_i} \simeq 1 -3.4~\frac{m_L}{\mu}.
  \label{eq:AM3}
\end{equation}
Both accretion of mass by the secondary star and the consideration
of the angular momentum of the spinning primary envelope will
reduce the orbital separation even faster.

\begin{figure}[h!]
\begin{center}
\includegraphics[width=80mm]{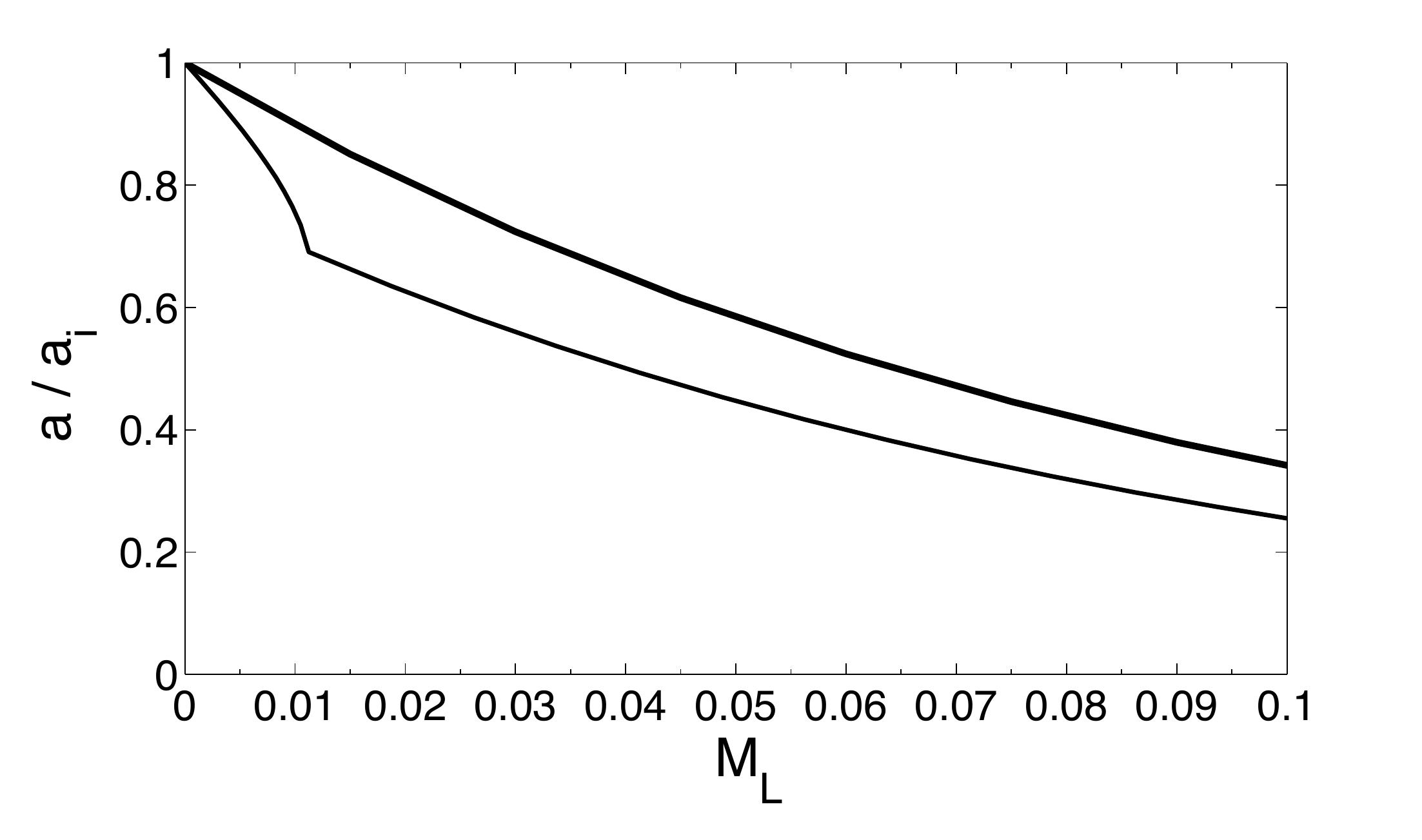}
\caption{The decrease in orbital separation, relative to the
initial separation, as function of mass lost from the second
lagrangian point $L_2$ in units of solar mass. The calculations
are for a primary star of $1.5 M_\odot$ and a secondary of $0.4
M_\odot$, hence $q=0.267$. The initial orbital separation is
$a_i=2 \AU$, the primary radius $R_1=1 \AU$ such that it just
fills its Roche lobe, the envelope mass is $M_{\rm env}=0.9
M_\odot$, and the moment of inertia of the giant primary star is
$I_1=0.22 M_{\rm env} R^2_1$. The upper line is for a case where
the change in the angular momentum of the spinning primary
envelope is neglected, according to equation (\ref{eq:AM3}). The
lower line is for a case where the primary stellar moment of
inertia is taken into account as long as the system is Darwin
stable (for $M_L<0.011M_\odot$ and $a>1.37 \AU$ for the parameters
used here), and the change in the angular momentum of the spinning
primary envelope is neglected afterward when the system becomes
Darwin unstable (at the break in the lower line: $M_L >
0.011M_\odot$).     }
 \label{fig:Asep}
\end{center}
\end{figure}

We numerically integrated equation (\ref{eq:AM1}) taking into
account the primary envelope moment of inertia. As the secondary
star spirals-in it spins-up the primary envelope. Synchronization
is maintained as long as the system is Darwin-stable. For
simplicity we neglect the spin-up of the primary envelope by the
secondary star after the system becomes Darwin unstable. In Fig.
\ref{fig:Asep} we present the results for one case with parameters
that might fit the progenitor of the Necklace PN. The initial
primary mass on the AGB is $1.5 M_\odot$ and the secondary mass is
$0.4 M_\odot$, hence $q=0.267$. The initial orbital separation is
$a_i=2 \AU$, and the primary radius is $R_1=1 \AU$ such that it
just fills its Roche lobe, the envelope mass is $M_{\rm env}=0.9
M_\odot$, and the moment of inertia of the giant primary star is
$I_1=0.22 M_{\rm env} R^2_1$. The system reaches Darwin
instability when the orbital separation decreases to $1.37 \AU$,
as seen by the `knee' in the lower line. The upper line is
according to equation (\ref{eq:AM3}), i.e., when the change in the
spin angular momentum of the primary star is neglected from the
beginning.

Equation (\ref{eq:AM3}) and Fig. \ref{fig:Asep} show that a small
mass loss through $L_2$, ${m_L} \la 0.1 {M_2}$ can bring the
system to form a CE. This, in principle, can account for the low
mass of the rings in the systems discussed in section
\ref{sec:intro}. However, those systems have also bipolar
structures, and some more mass might be lost from the system
during this phase. As well, the secondary star might accrete some
mass. Over all, very little mass will be ejected through $L_2$
during the Darwin unstable phase, ${m_L} < 0.1 {M_2}$. Although
this is compatible with observations of these objects, the energy
of the gas at $L_2$ poses some problems to this interpretation, as
we now show.

\subsection{Energy at $L_2$}
\label{subsec:EL2}

\begin{figure}[h!]
\begin{center}
\includegraphics[width=60mm]{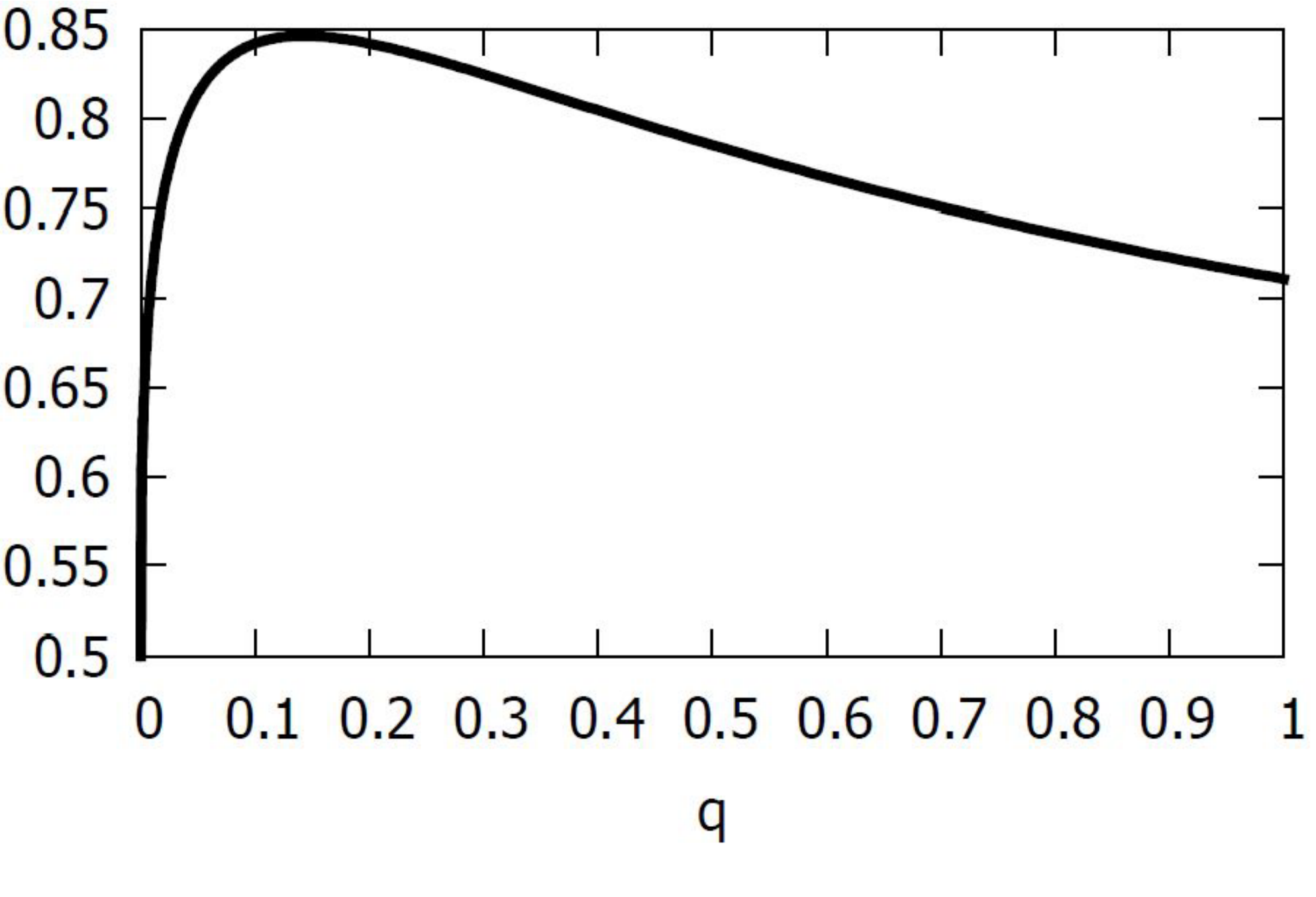}
\caption{The ratio of kinetic energy to the value of the
gravitational energy at $L_2$. As this ratio is $<1$, if no
further acceleration takes place the matter at $L_2$ stays bound
to the binary system. }
 \label{fig:EkinEpot}
\end{center}
\end{figure}

We calculated the specific kinetic energy of the mass leaving at
$L_2$ assuming it has the velocity of the point $L_2$. The ratio
of the kinetic energy relative to the value of the gravitational
energy at $L_2$ is presented in Fig. \ref{fig:EkinEpot}. For all
value of mass ratio $q$ this ratio is below 1, implying that the
mass leaving $L_2$ stays bound to the binary system if no further
outward acceleration takes place.
The final energy will be larger because of the torque exerted on
the gas after it leaves $L_2$ \citep{Lin1977}. Still, the terminal
velocity is smaller than the Keplerian velocity at $L_2$.

The expansion velocity of the equatorial ring in the Necklace is
$28 \km \s^{-1}$, larger than the typical escape velocity from AGB
stars and from the typical velocity of AGB winds. In light of the
results presented in Fig. \ref{fig:EkinEpot}, the expansion
velocity of the ring in the Necklace calls for one of three
explanations. (1) The ring was expelled at the end of the CE
phase, during the so called migration phase where a circumbinary
gas is presence \citep{Soker2013, Soker2014}. Then the orbital
separation is $\sim 10 R_\odot$, and the escape velocity from the
binary system is $\sim 200 \km \s^{-1}$. The expansion velocity of
the ring is a small fraction of that velocity. This explanation
for ring formation is not applicable to the equatorial ring of
SN~1987A, as the binary system there did not terminate a CE phase,
and hence we will not explore this possibility here. (2) If the
ring was expelled during the AGB phase, it must have experienced
an extra acceleration after it left $L_2$, much more than the
torque by the binary system. (3) The ring was formed when the hot
AGB remnant experienced an outburst, e.g., a born-gain AGB
process. This explanation holds for the formation of the rings in
the born-again PNe A30 and A78 that were discussed in section
\ref{sec:intro} and for novae. We think that the Necklace was
formed by option (3), with jets launched by the secondary star.

In principle, the matter leaving $L_2$ during an AGB phase could
be accelerated by radiation from the giant star or by the thermal
pressure of the gas. We see a problem with such a process if a
narrow ring is to be formed at later times. The thermal pressure,
either from the gas at $L_2$ being hot or from radiation pressure,
will cause the equatorial gas to expand not only in the equatorial
plane, but also in the directions perpendicular to the equatorial
plane. It is not clear whether a narrow ring can be formed from
such a process. We turn to examine acceleration and focusing of
the equatorial outflow by jets.

\section{NUMERICAL SIMULATIONS OF JETS}
\label{sec:hydro}

 Motivated by the presence of bipolar structures in many (or all)
of the systems with equatorial rings (section \ref{sec:intro}),
and by the energetic problem discussed in section
\ref{subsec:EL2}, we turn to study the formation of a ring by
bipolar jet pairs, following the mechanism proposed by
\cite{SokerRappaport2000}. In many systems the main bipolar
structure was formed before or after the equatorial ring. Non the
less, we assume that weaker bipolar jets (or a collimated fast
wind; CFW) are launched during the ring formation phase. This can
occur either during a Darwin unstable phase or during a stable
phase. The same mass that fills the gravitational-potential lobe around the
secondary that crosses $L_2$, is accreted in part onto the
secondary star through an accretion disk. The secondary then is
assumed to launch two opposite jets.

\subsection{Numerical setup}
\label{subsec:setup}

Our simulations are performed by using version 4.0-beta of the
FLASH code \citep{Fryxell2000}. The FLASH code is an adaptive-mesh
refinement (AMR) modular code used for solving hydrodynamics or
magnetohydrodynamics problems. Here we use the unsplit PPM
(piecewise-parabolic method) solver of FLASH. We neither include
gravity, as velocities are much above the escape speed in the
region we simulate, nor radiative cooling,  as the interaction
region is optically thick. Instead of calculating radiative
cooling and radiative transfer, that are too complicated for the
flow geometry, we simulate two values of the adiabatic index
$\gamma$, $4/3$ and $5/3$. Using lower values of $\gamma<5/3$ to
mimic radiative cooling is reasonable when kinetic energy is
channelled to thermal energy, but not when thermal energy is
channelled to kinetic energy. For more details on the numerical
settings and the justifications for the parameters employed in
this study see \cite{AkashiSoker2013}.

We employ a full 3D AMR (9 levels; $2^{12}$ cells in each
direction) using a Cartesian grid $(x,y,z)$ with outflow boundary
conditions at all boundary surfaces. We take the $x-y$ plane with
$z=5\times10^{15} \cm$ to be in the equatorial plane of the PN,
and we simulate the whole space (the two sides of the equatorial
plane).

At $t=0$ we place a spherical dense shell in the region $R_{\rm
in} = 10^{14} \cm < r < 2\times 10^{14} \cm= R_{\rm out}$, and
with a density profile of $\rho_s = 1.58 \times 10^{-11}(r/10^{14}
\cm)^{-2} \g \cm^{-3}$, such that the total mass in the shell is
$0.1M_\odot$. The gas in the shell has an initial radial velocity
of $v_s = 10 \km \s^{-1}$. The shell corresponds to a mass loss
episode lasting for $\sim 3 \yr$ and with a constant mass loss
rate of $\dot M_s \simeq 0.03 M_\odot \yr^{-1}$.  Such an event
can be classified as an ILOT, as described in
\cite{AkashiSoker2013}, where more details can be found. The
regions outside and inside the dense shell are filled with a lower
density spherically-symmetric slow wind having a uniform radial
velocity of  $v_{\rm wind}=v_s= 10 \km \s^{-1}$. The slow wind
density at $t=0$ is taken to be $\rho(t=0) = {\dot M_{\rm
wind}}({4 \pi r^{2} v})^{-1}$, where $\dot M_{\rm wind}=10^{-5}
{\rm M_\odot \yr^{-1}}$.

The two opposite jets are lunched from the inner $10^{14} \cm $
region along the $z$-axis and  within a half opening angle of
$\alpha = 50^\circ$. By the term `jets' we refer also to wide
outflows, as we simulate here. More generally, we simulate
slow-massive-wide (SMW) outflows.  Although the jets that are
simulated here are not observed (because the medium is optically
thick), such wide outflows are commonly observed from active
galactic nuclei (e.g., \citealt{Aravetal2013}). This, and the
success of wide outflows to explain lobes observed in cooling flow
clusters (e.g., \citealt{Sternbergetal2007}) motivate us to
consider wide outflows. 

The launching episode lasts for $5 \times 10^{6} \s = 58 \days$.
The jets' initial velocity is $v_{\rm jet}=1000 \km \s^{-1}$, and
the mass loss rate into the
two jets together is $\dot M_{\rm 2jets} = 0.13 M_\odot \yr^{-1}$. 
The slow wind, dense shell, and the ejected jets start with a
temperature of $1000 \K$. The initial jets' temperature has no
influence on the results (as long it is highly supersonic) because
the jets rapidly cool due to adiabatic expansion. For numerical
reasons a weak slow wind is injected in the sector
$\alpha<\theta<90^\circ$ (more details are in
\citealt{AkashiSoker2013}).

\subsection{Results of main model}
\label{subsec:results}

 To describe the flow structure
of our 3D simulations we will present the density, pressure, and
velocity maps in the meridional plane, i.e. a plane through the
symmetry axis of the jets, and in planes parallel to the
equatorial plane.

\begin{figure}
\subfigure[$t=0$~days]{\includegraphics[height=2.8 in,width=3.in,angle=0]{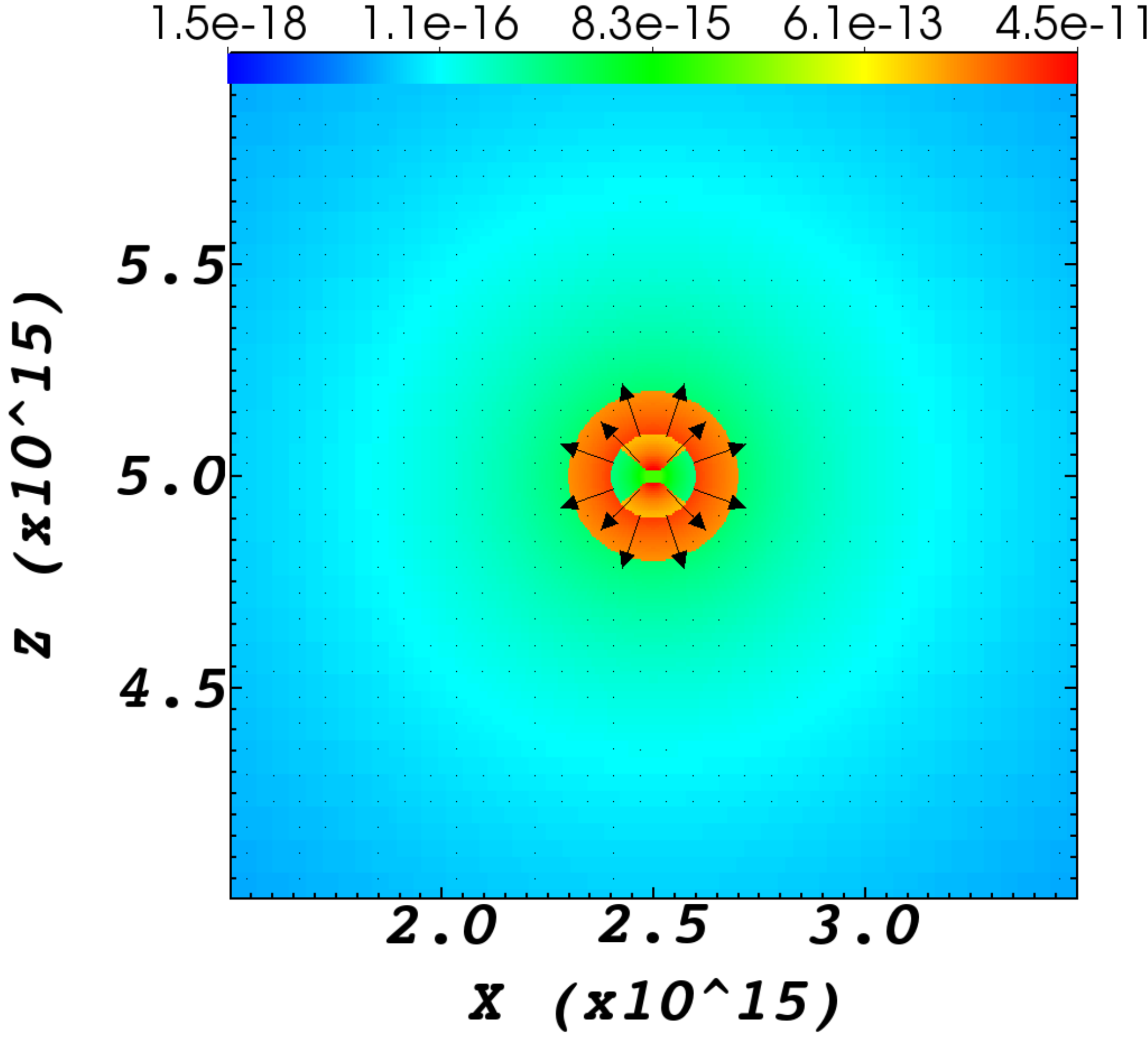}}
\hskip -0.1cm
\subfigure[$t=47$~days]{\includegraphics[height=2.8 in,width=3.in,angle=0]{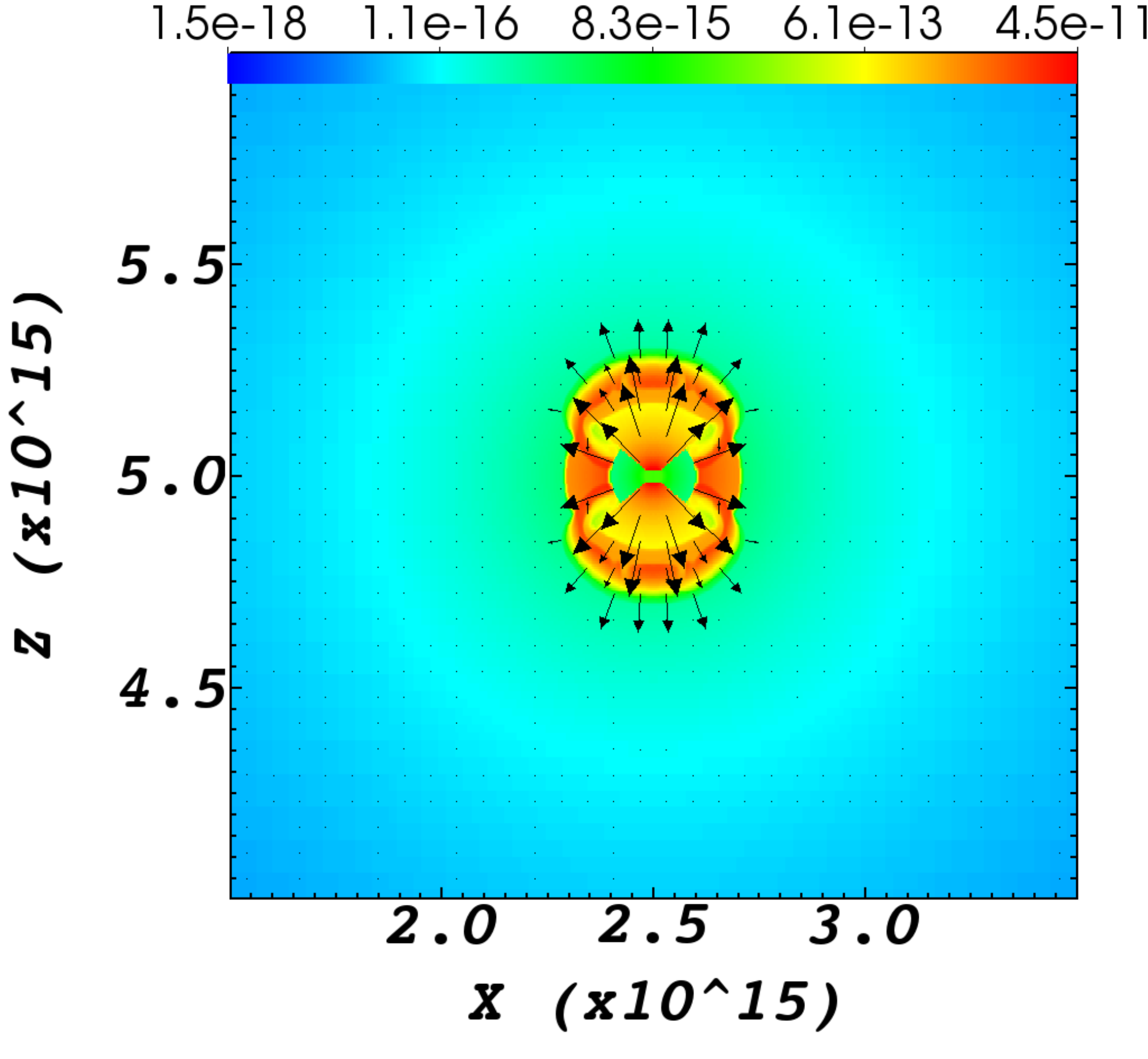}}
\vskip +.2cm
\subfigure[$t=94$~days]{\includegraphics[height=2.8in,width=3.in,angle=0]{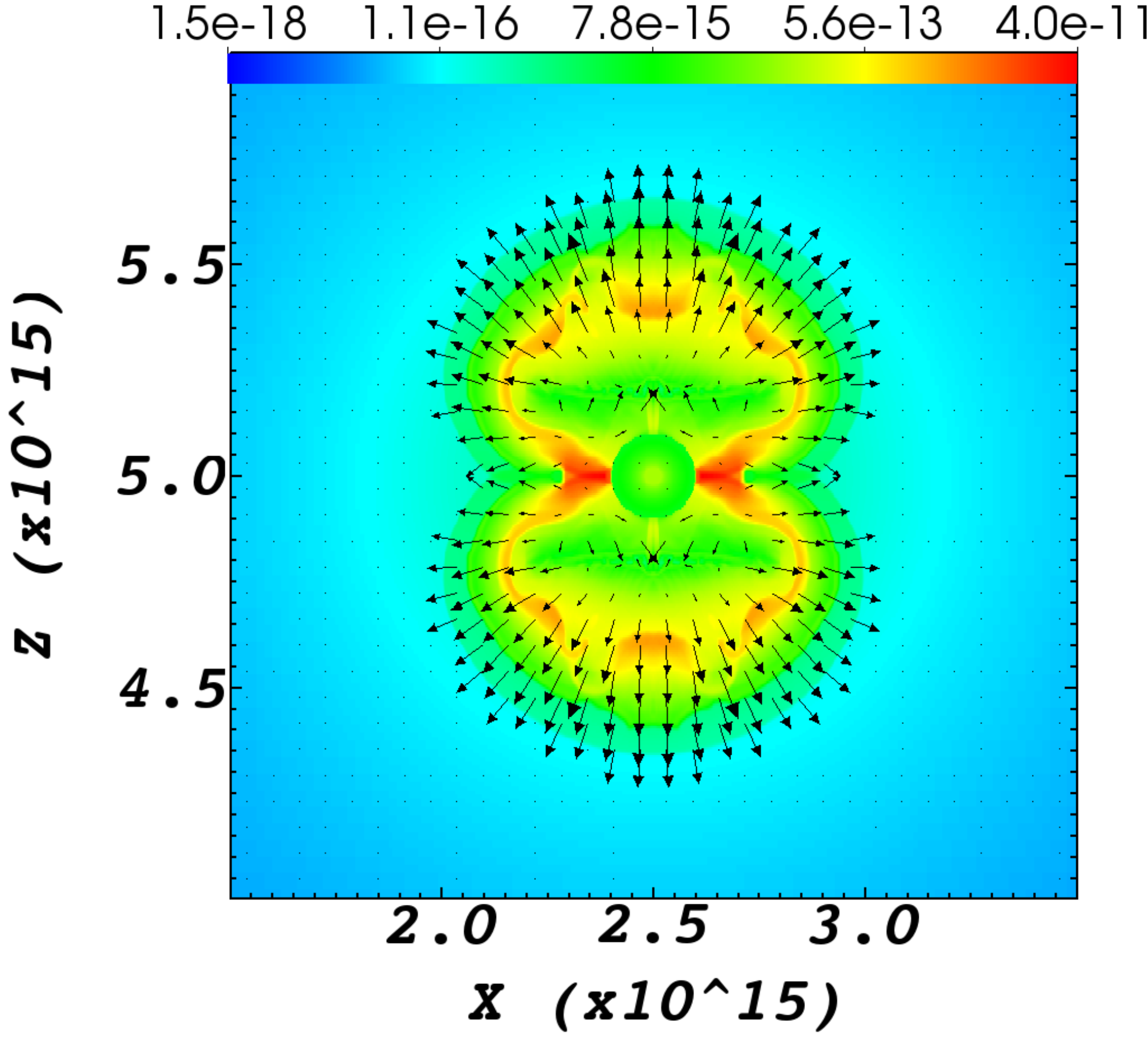}}
\hskip -0.15cm
\subfigure[$t=140$~days]{\includegraphics[height=2.85in,width=3.25in,angle=0]{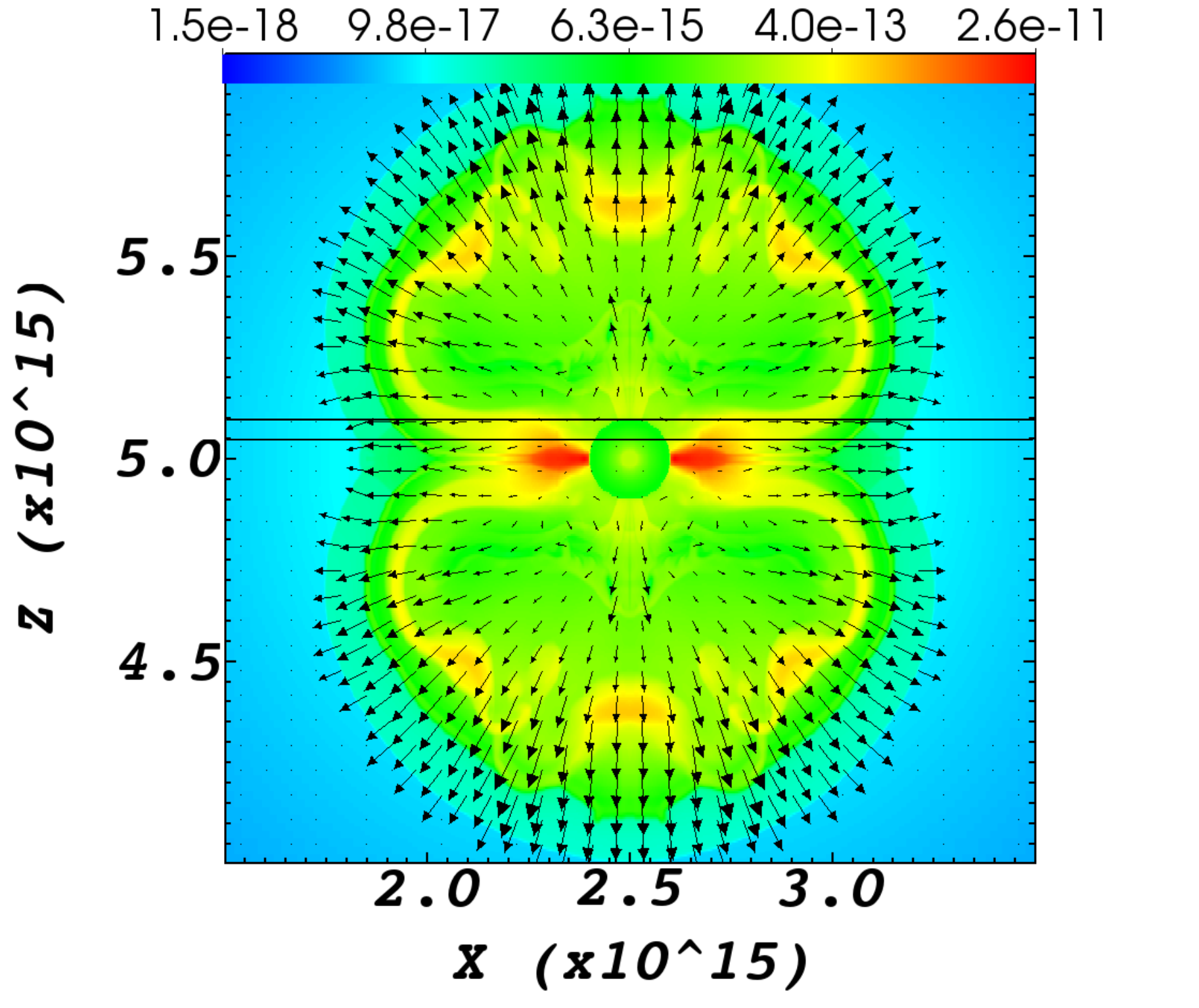}}
\caption{The density maps in the meridional plane  at four
times for the $\gamma=5/3$ run. Two opposite jets are injected at
$r=10^{14} \cm$ along the Z axis, with a half opening angle of
$50^\circ$, with an initial velocity of $1000 \km \s^{-1}$, and
during the time period $t=0$ to $t=58~$days. The shell, seen in
upper left panel, and the low density gas filling the rest of the
volume start with a radial velocity of $10 \km \s^{-1}$. Arrows
depict the flow direction and the velocity magnitude that is
relative to the arrow length; the arrows near the center in the
first two panels are for $1000 \km \s^{-1}$. Density color coding
is in units of $\g \cm^{-3}$. The times of panels a, b, c, and d
are $0$, $47$, $94$, and $140$ days, respectively. Units on the
axes are in $\cm$. The two horizontal lines in panel (d) show the
planes presented in Fig. \ref{fig:g4}. Note the development of a
dense region in and near the equatorial plane that flows parallel
to the equatorial plane. We propose that this equatorial flow
turns to a dense ring later on.}
 \label{fig:g3}
\end{figure}

We start with the adiabatic index of  $\gamma=5/3$ case. In Fig.
\ref{fig:g3} we present the time evolution of the density in the
meridional plane at four times. The jets were injected at
$r=10^{14} \cm$ along the Z axis, with a half opening angle of
$50^\circ$, and with an initial velocity of $1000 \km \s^{-1}$.
The jets are active in the time period $t=0$ to $t=58~$days. The
regions outside the shells are filled with lower density gas with
a radial velocity of $10 \km \s^{-1}$. The arrows on the density
maps present the velocity direction and magnitude at each point.

\begin{figure}
\subfigure[$t=47$~days]{\includegraphics[height=2.8in,width=3.65in,angle=0]{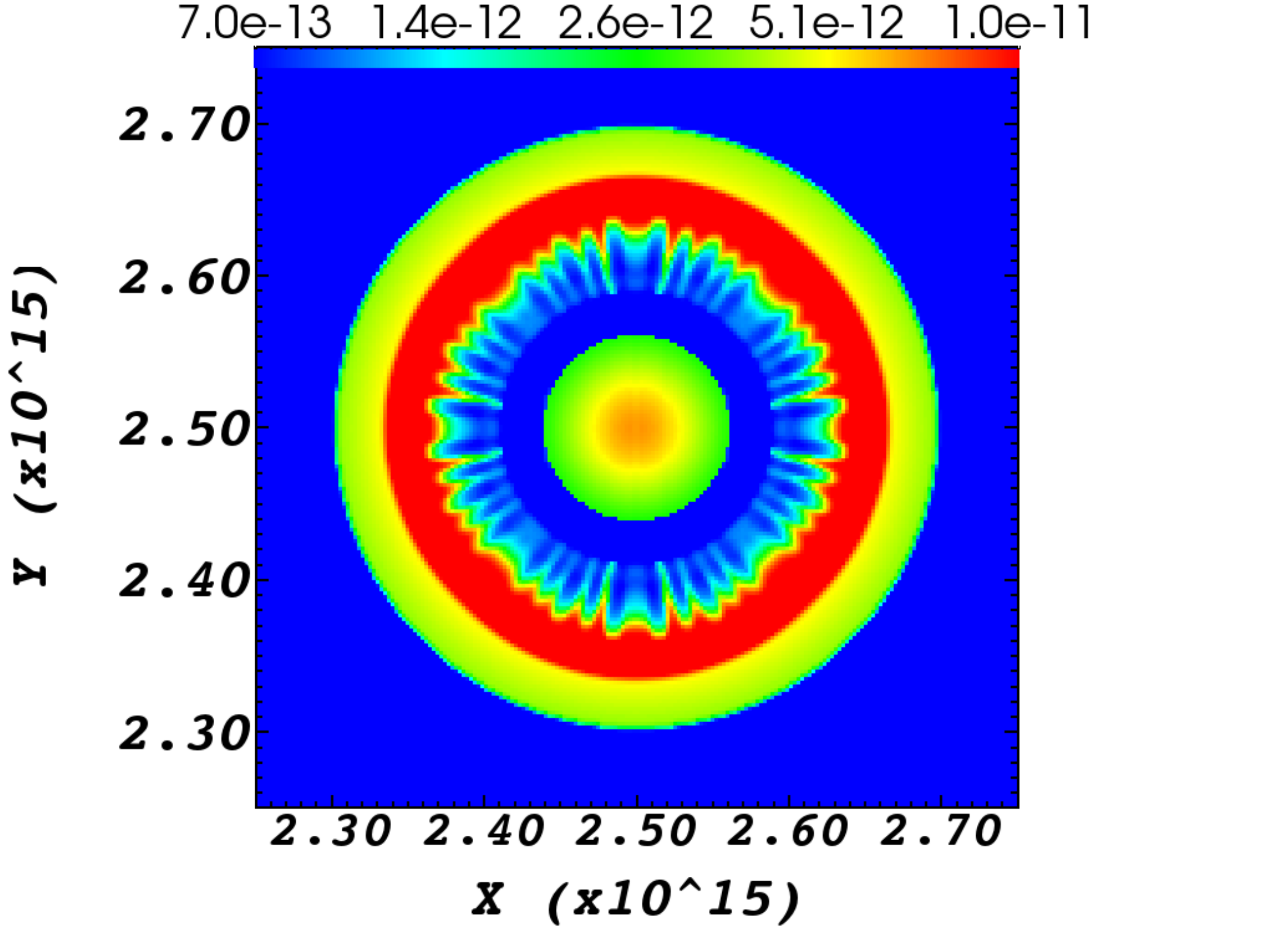}}
\hskip -0.7cm
\subfigure[$t=140$~days]{\includegraphics[height=2.8in,width=3.65in,angle=0]{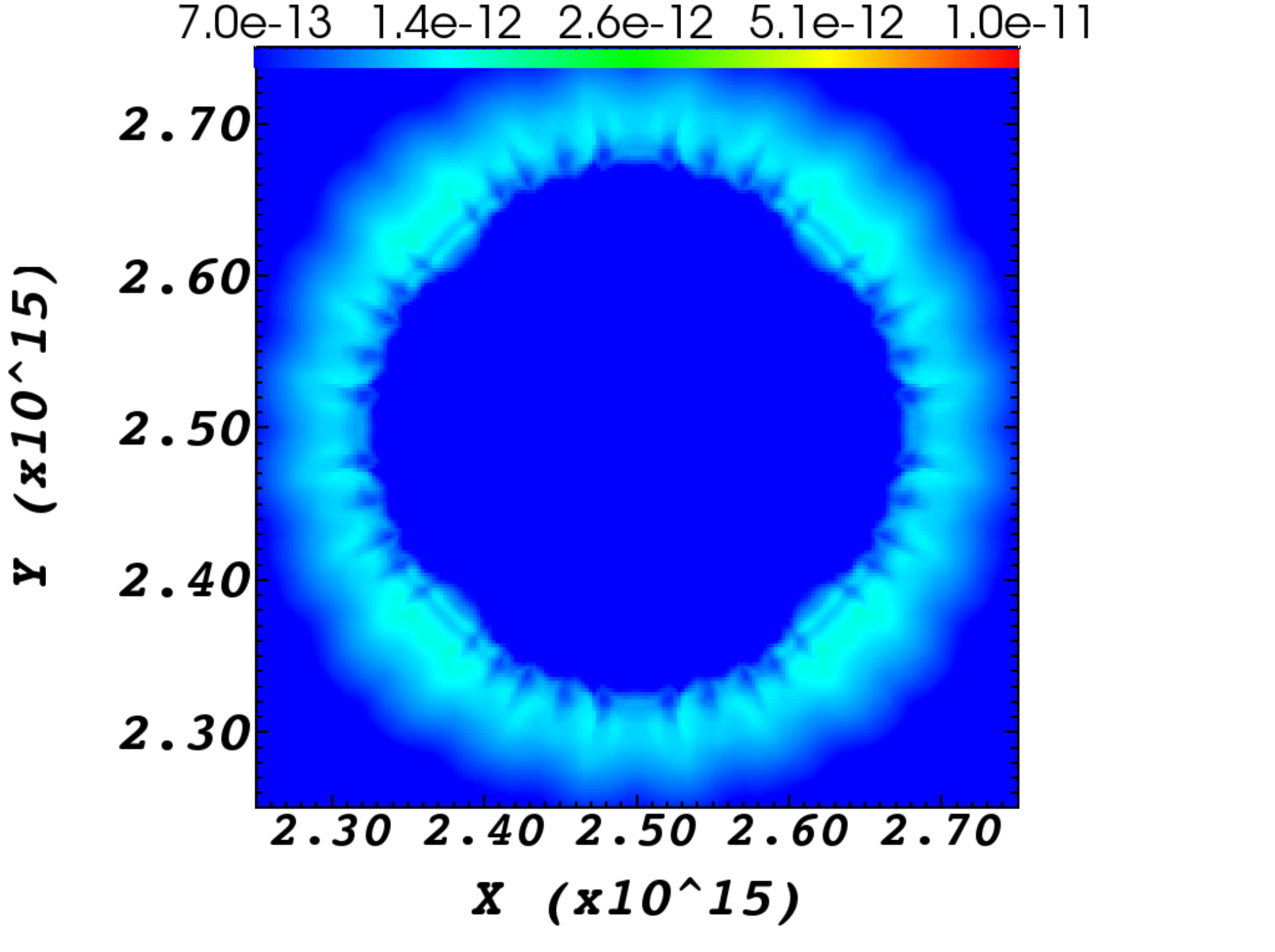}}
\subfigure[$t=47$~days]{\includegraphics[height=2.8in,width=3.65in,angle=0]{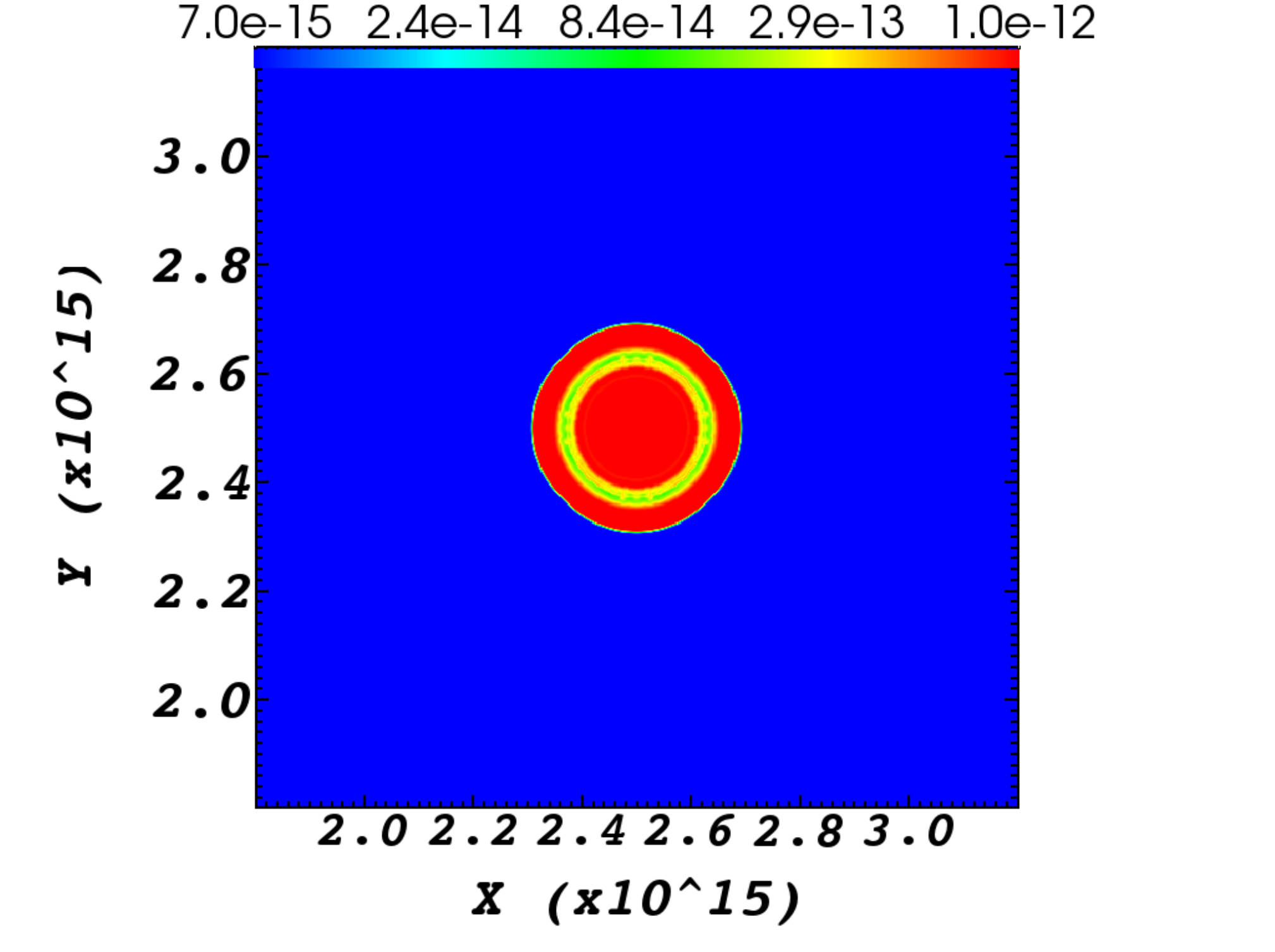}}
\hskip -0.7cm
\subfigure[$t=140$~days]{\includegraphics[height=2.8in,width=3.65in,angle=0]{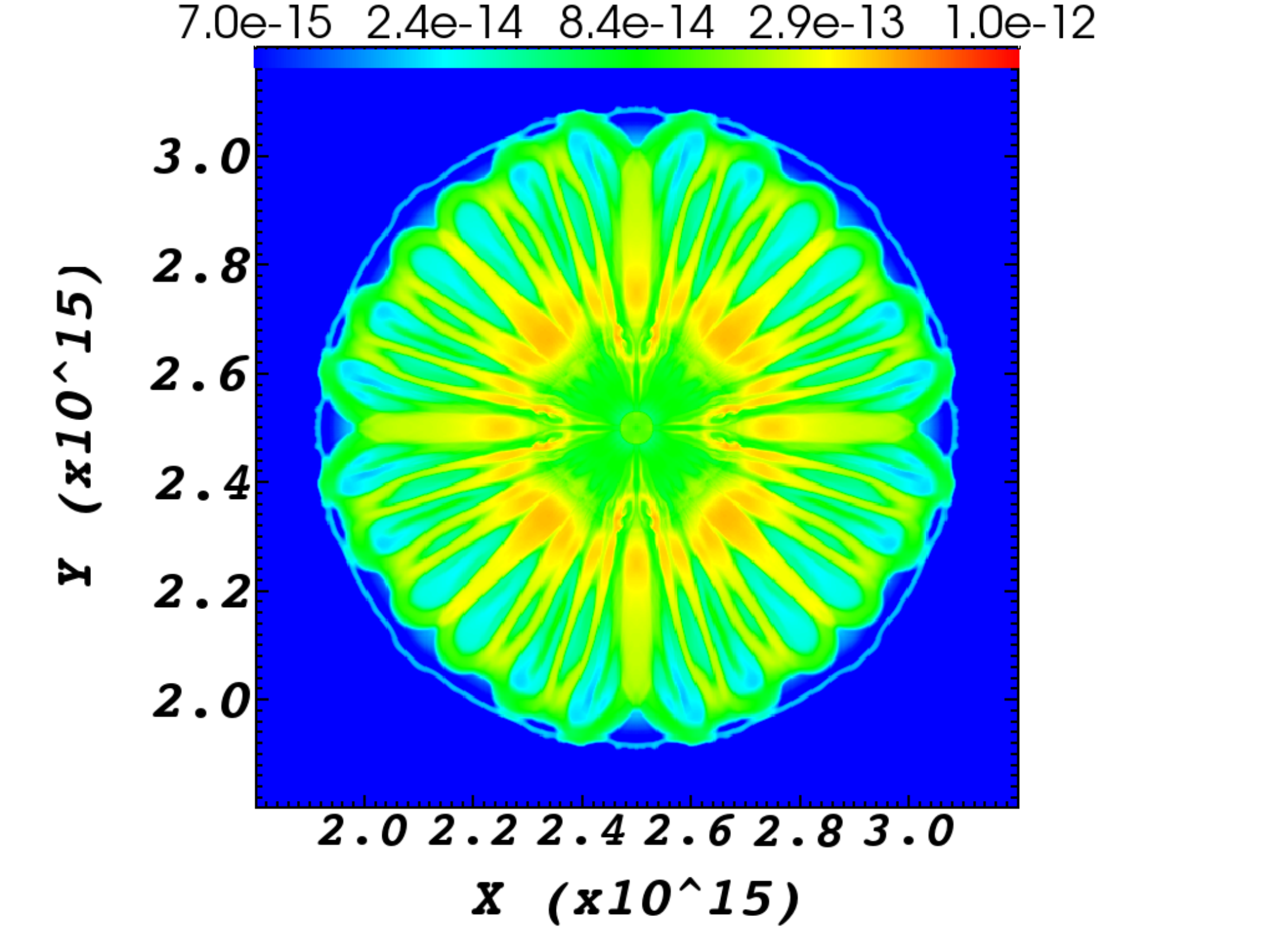}}
\caption{The density maps  at four times of the
$\gamma=5/3$ run, and parallel to the equatorial plane. Color
coding is in $\g \cm^{-3}$. Left panels are given at times 47 days
and right panels at 140 days. Upper panels are the plane $z=5.05
\times 10^{15} \cm$, i.e. $\Delta z= 0.05 \times 10^{15} \cm$ from
the equatorial plane, and the lower panels are for $z=5.1 \times
10^{15} \cm$, i.e. $\Delta z= 0.1\times 10^{15} \cm$. Note the
development of instabilities at the later times in both planes.
These will later form a clumpy ring. The exact structure and
location of the instabilities are determined also by the numerical
gird, but their existence is physical. Units on the axes are in
$\cm$. } \label{fig:g4}
\end{figure}

In Fig. \ref{fig:g4}  we present the density maps at two times, 47
days on the left panels and 140 days on the right panels, and in
two planes parallel to the equatorial plane. Upper panels are the
plane $z=5.05 \times 10^{15} \cm$, i.e. $\Delta z= 0.05 \times
10^{15} \cm$  from the equatorial plane, and the lower panels are
for $z=5.1 \times 10^{15} \cm$, i.e. $\Delta z= 0.1\times 10^{15}
\cm$. These two planes are marked by horizontal lines in the
$t=140$~days panel of Fig. \ref{fig:g3}.

\begin{figure}
\subfigure[$t=140$~days]{\includegraphics[height=3.8in,width=4.65in,angle=0]{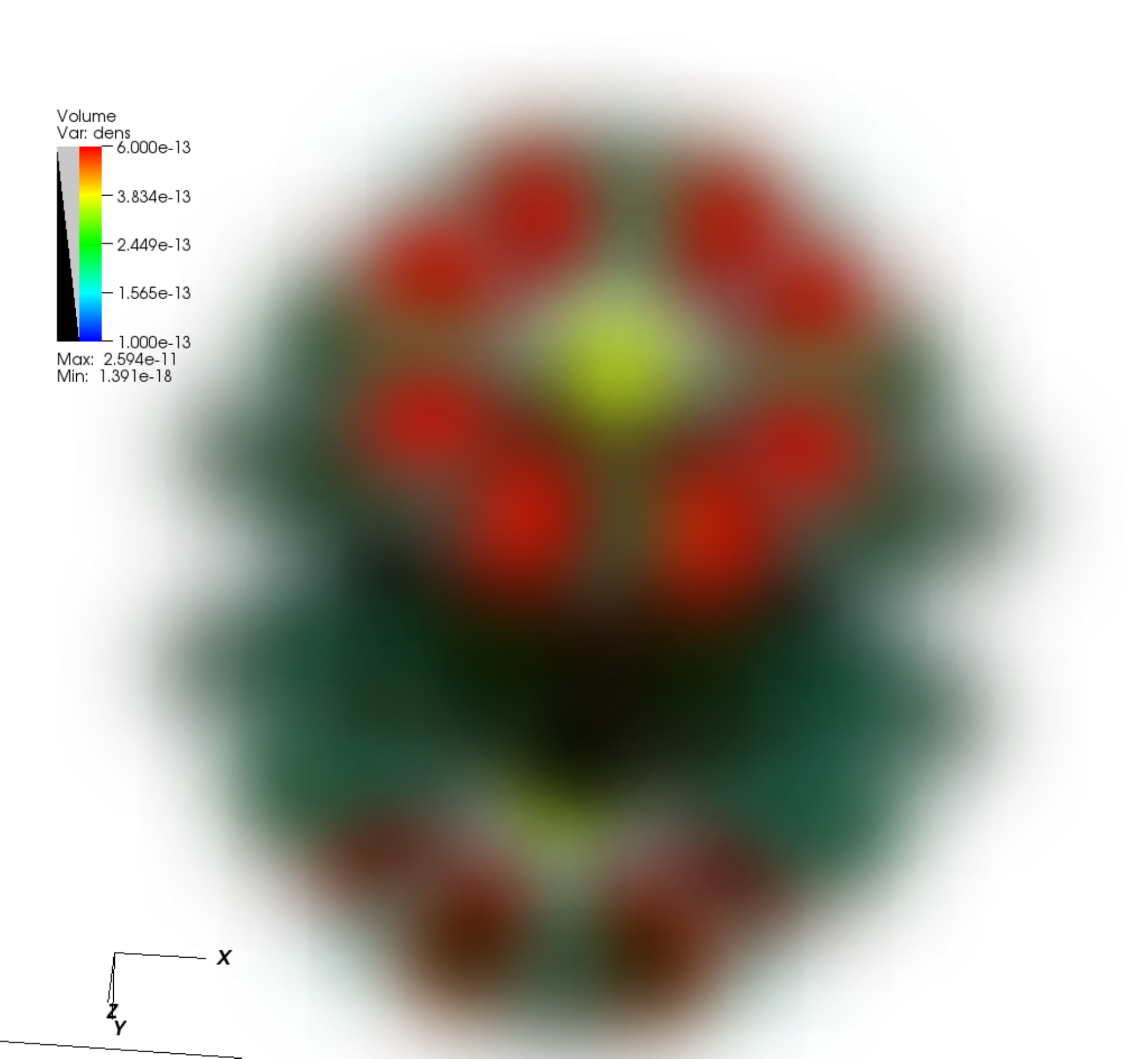}}
\hskip -0.7cm
\subfigure[$t=140$~days]{\includegraphics[height=3.8in,width=3.35in,angle=0]{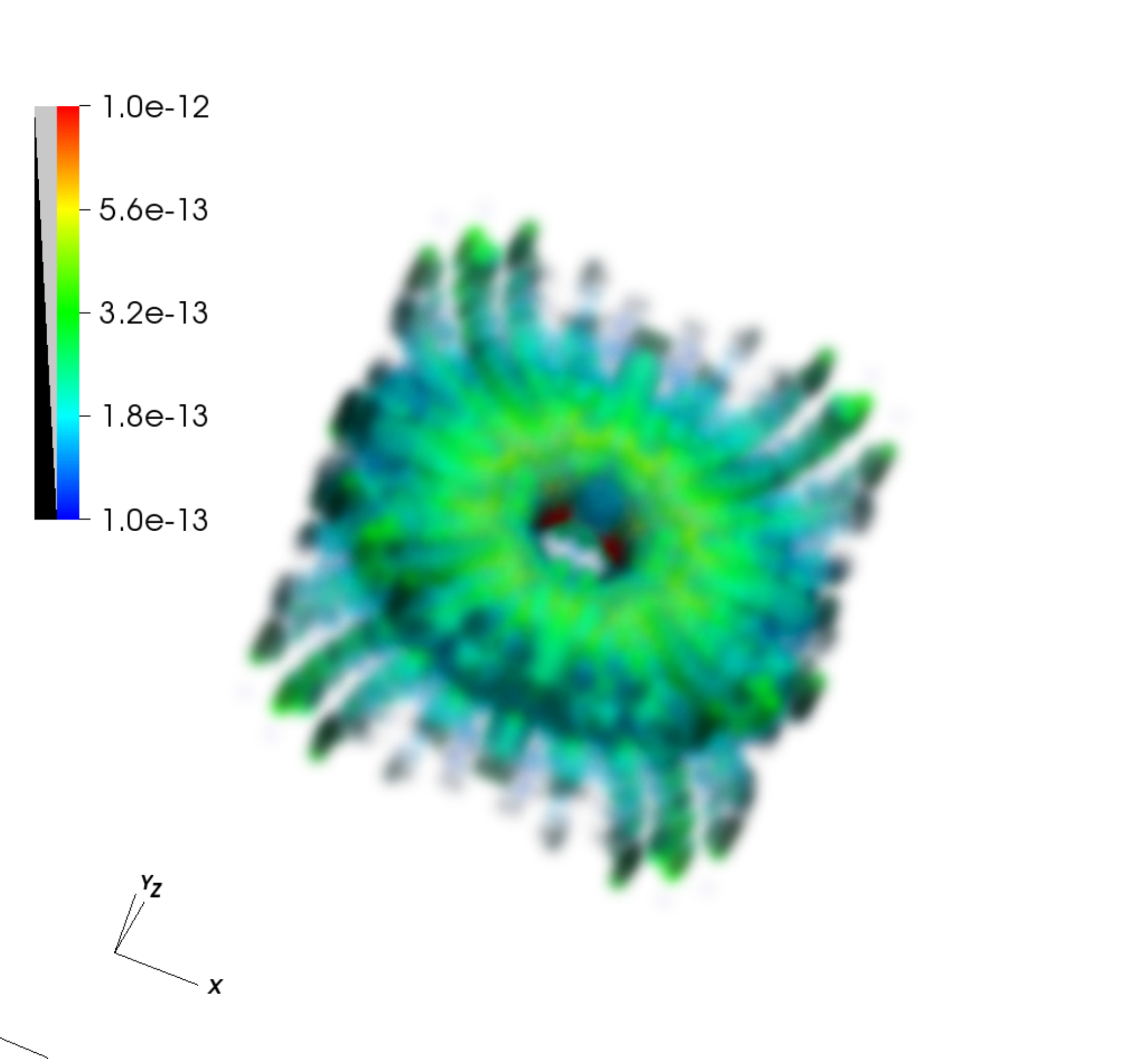}}

\caption{Three dimensional density structure for the $\gamma=5/3$ run (main run)
presented in figures \ref{fig:g3} and \ref{fig:g4}.
In the left panel, two clumpy rings, the red clumps, are seen in the upper and the lower parts of the nebula.
A single ring is seen near the equatorial plane in black.
In the right panel, the inner clumpy disk near the center is shown. The inner part of this ring is the densest part of the nebula (the red regions inside the ring). This equatorial ring is the focus of our study.
Density color coding is in units of $\g \cm^{-3}$. }
\label{fig:3D}
\end{figure}

We note the following properties of the flow for the $\gamma=5/3$
case. (1) The jets inflate two opposite lobes. Such a general
bipolar structure is observed in many PNe and were simulated
before (e.g., \citealt{AkashiSoker2008, sokeretall2013} ). (2) The
jets compress some gas toward the equatorial plane
(\citealt{AkashiSoker2008}). As seen at 94 and 140 days in figure
\ref{fig:g3}, a dense and slow equatorial flow is develops. This
dense flow will become a ring at late times. (3) The flow becomes
Rayleigh-Taylor (RT) unstable (see also
\citealt{AkashiSoker2013}). Large RT-tongues develop in the
meridional plane, as can be seen on each side of the symmetry axis
at times 94 and 140 days in Fig. \ref{fig:g3}.   (4) Other RT
unstable regions are seen in the right panels of Fig.
\ref{fig:g4}. These two panels are planes close to the equatorial
plane, along which the flow is parallel to the equatorial plane.
These many RT-tongues will lead to the formation of dense clumps
in the equatorial ring.  For this last case we also present in
Fig.~\ref{fig:3D} a three dimensional view of the density
structure. The two panels present the density surfaces for
different values of the density, as given in the color-bar, and as
explained in the caption.  Clearly, a clumpy slowly expanding
equatorial ring is formed by this flow structure. We note that the
exact structure and location of the instabilities are determined
also by the numerical grid structure, but their existence is
physical.

\begin{figure}
\subfigure[$t=0$~days]{\includegraphics[height=2.8in,width=3.65in,angle=0]{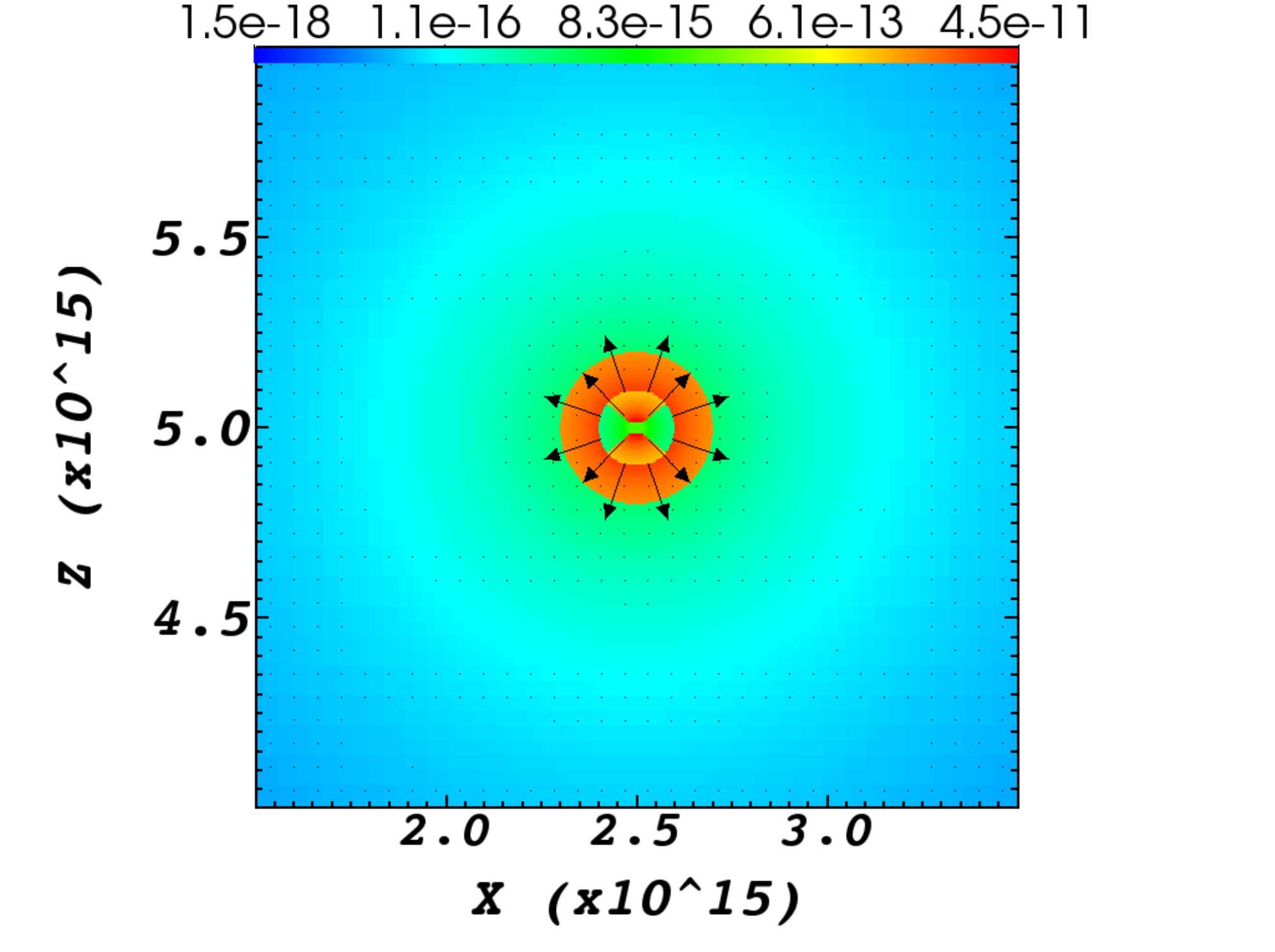}}
 \hfill
\hskip -0.7cm
\subfigure[$t=47$~days]{\includegraphics[height=2.8in,width=3.65in,angle=0]{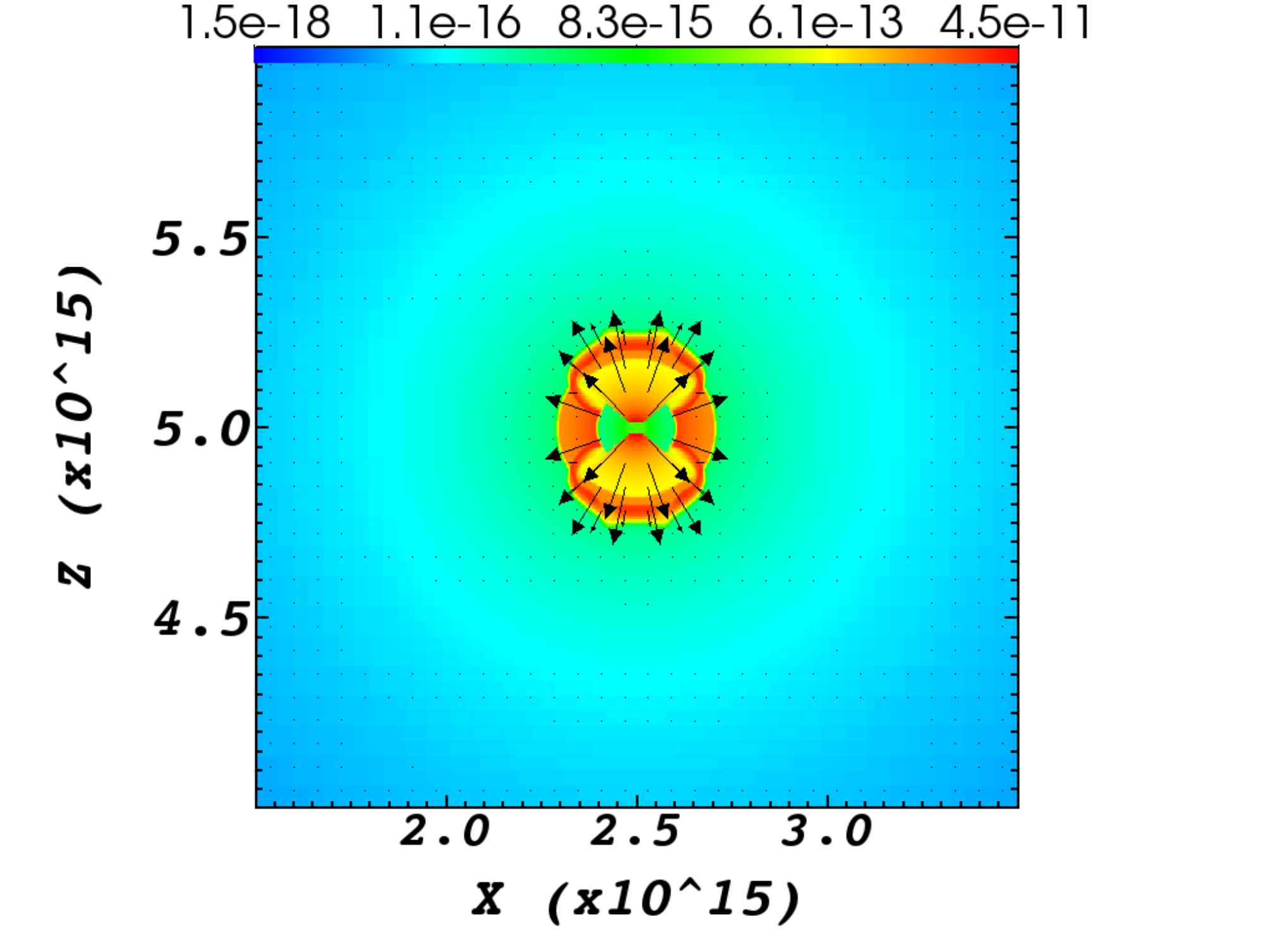}}
 \hfill
\subfigure[$t=94$~days]{\includegraphics[height=2.8in,width=3.65in,angle=0]{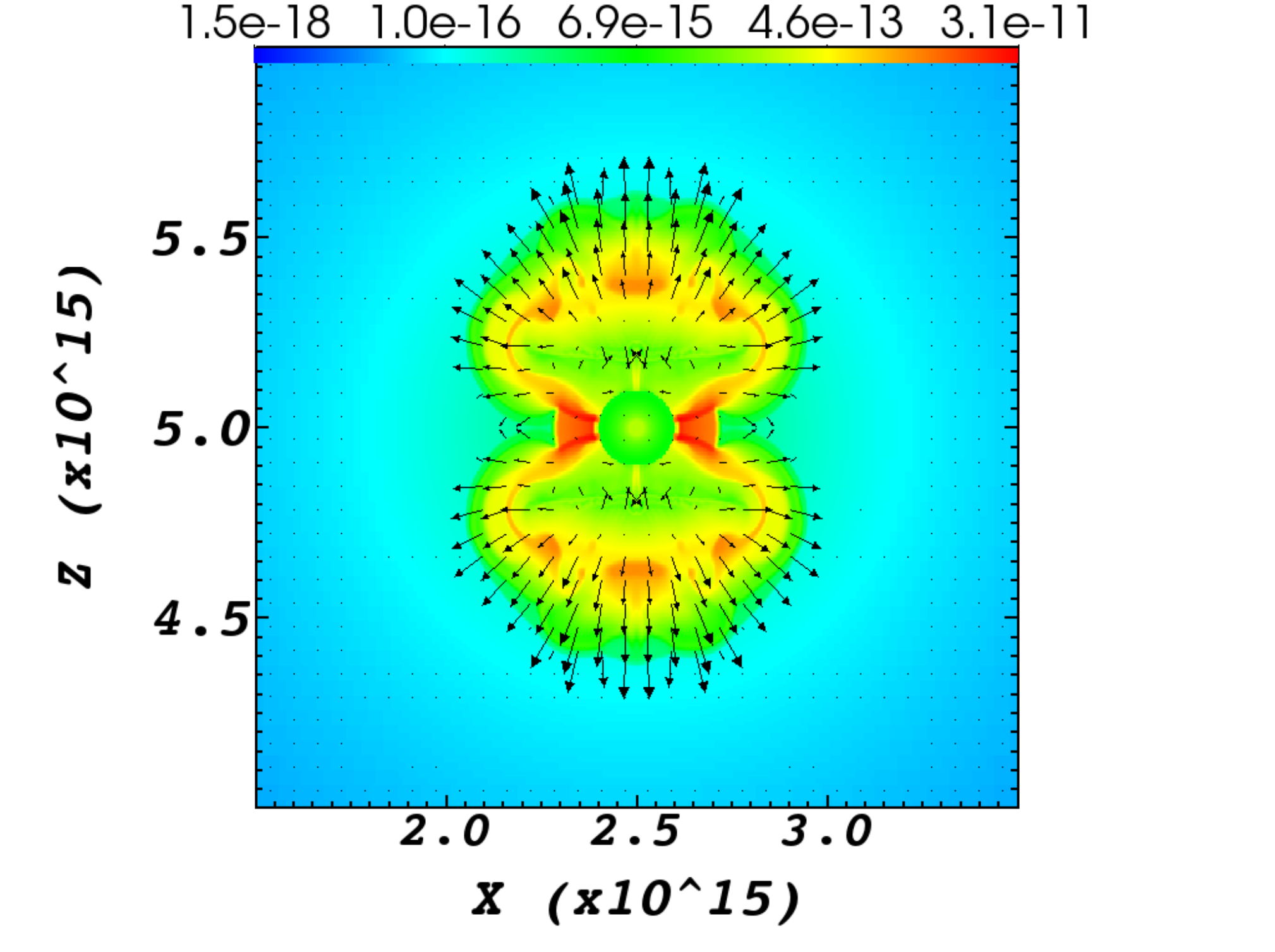}}
 \hfill
\hskip -0.7cm
\subfigure[$t=140$~days]{\includegraphics[height=2.8in,width=3.65in,angle=0]{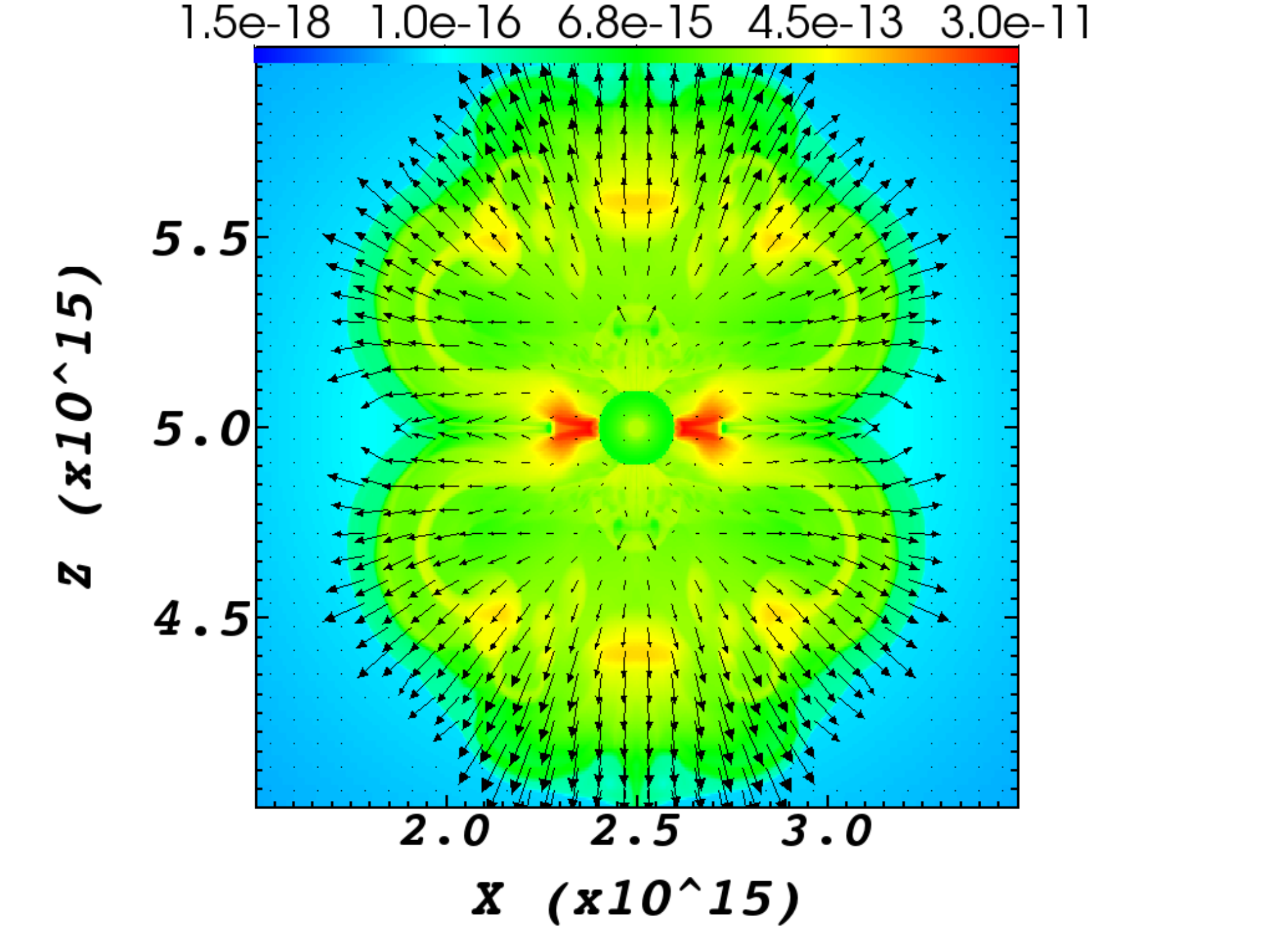}}
 \hfill
\caption{Like Fig. \ref{fig:g3} but for the $\gamma=4/3$
case. }
\label{fig:g5}
\end{figure}

\begin{figure}
\subfigure[$t=47$~days]{\includegraphics[height=2.8in,width=3.65in,angle=0]{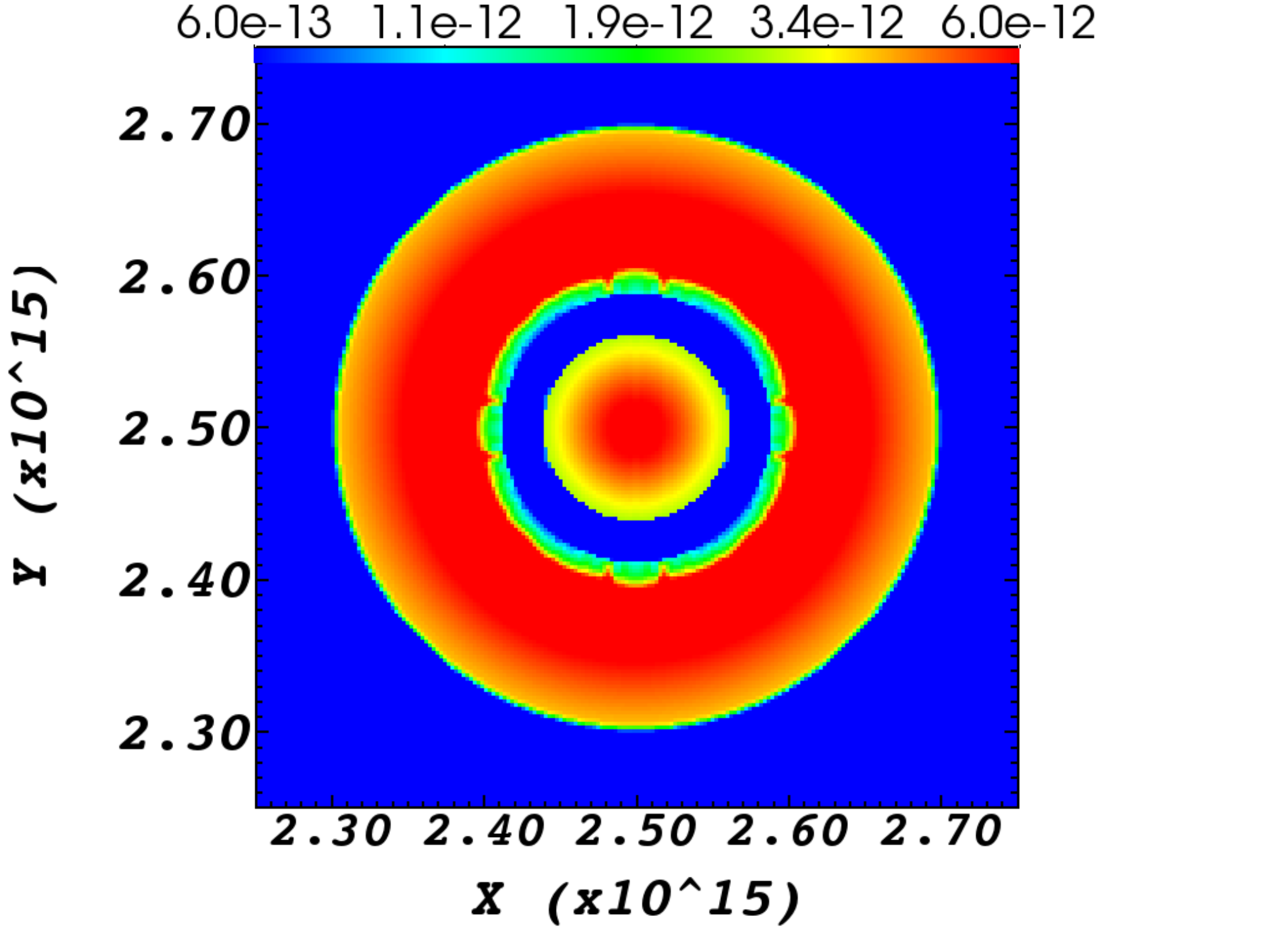}}
 \hfill
\hskip -0.7cm
\subfigure[$t=140$~days]{\includegraphics[height=2.8in,width=3.65in,angle=0]{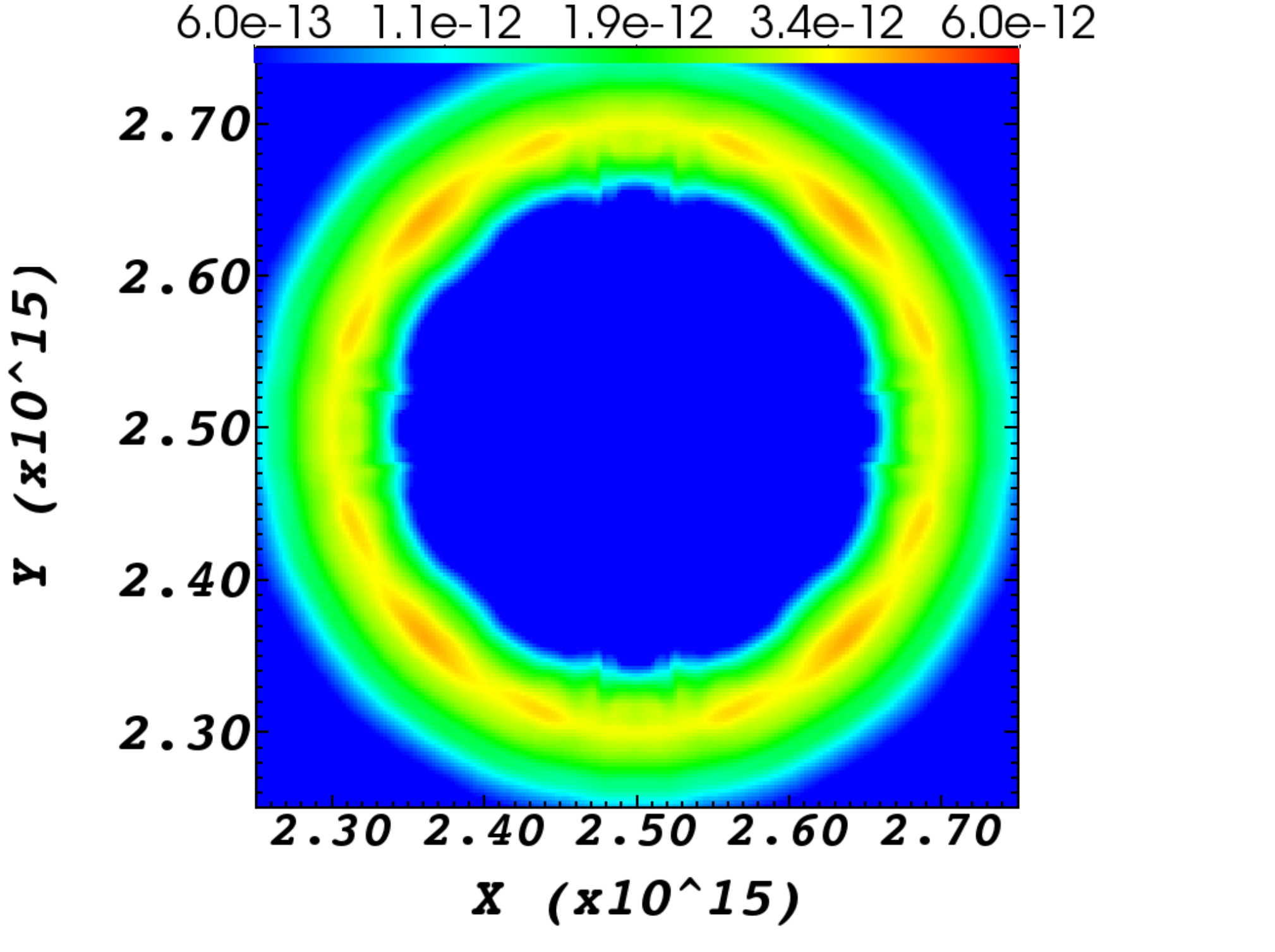}}
 \hfill
\subfigure[$t=47$~days]{\includegraphics[height=2.8in,width=3.65in,angle=0]{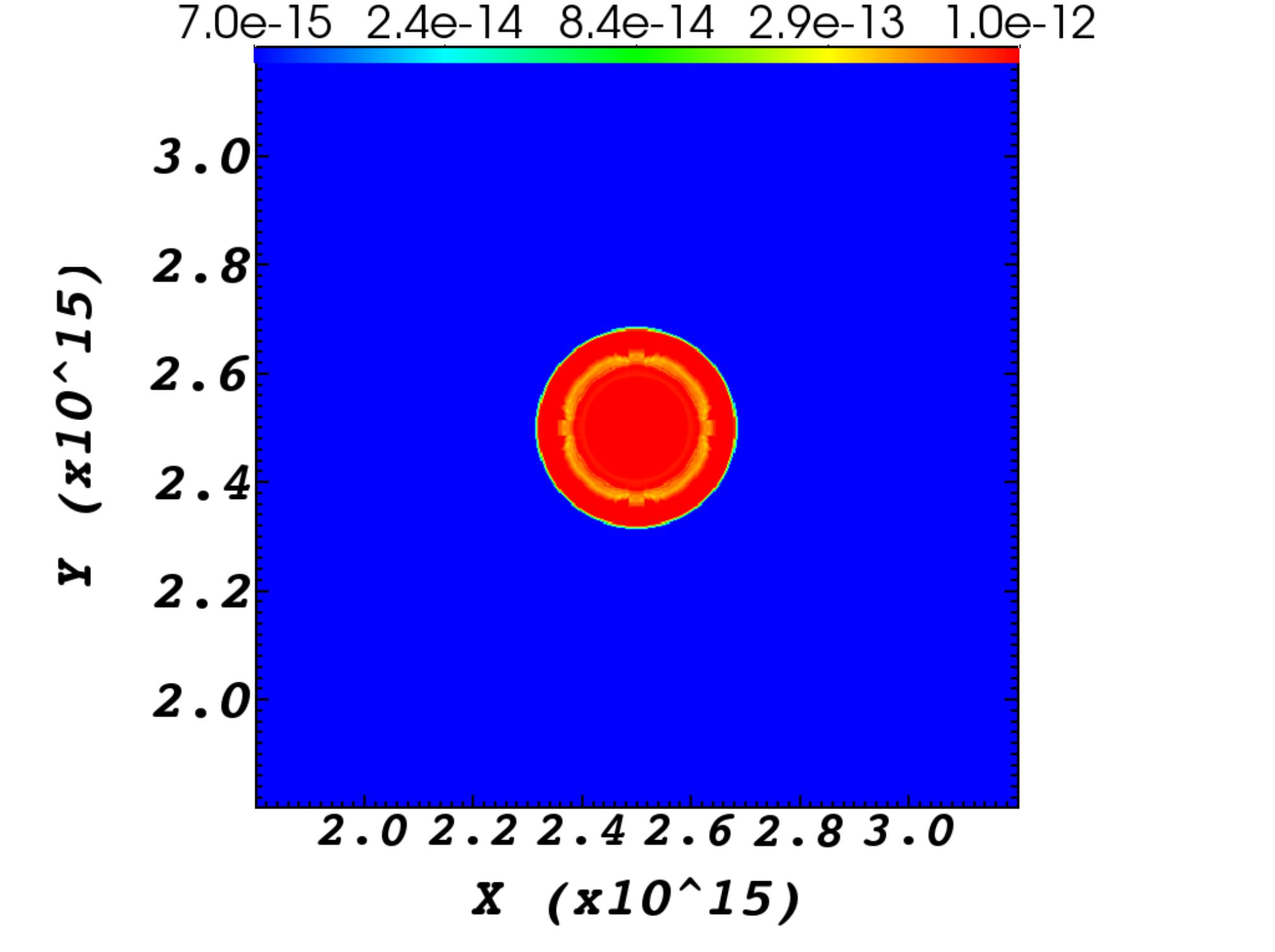}}
 \hfill
\hskip -0.7cm
\subfigure[$t=140$~days]{\includegraphics[height=2.8in,width=3.65in,angle=0]{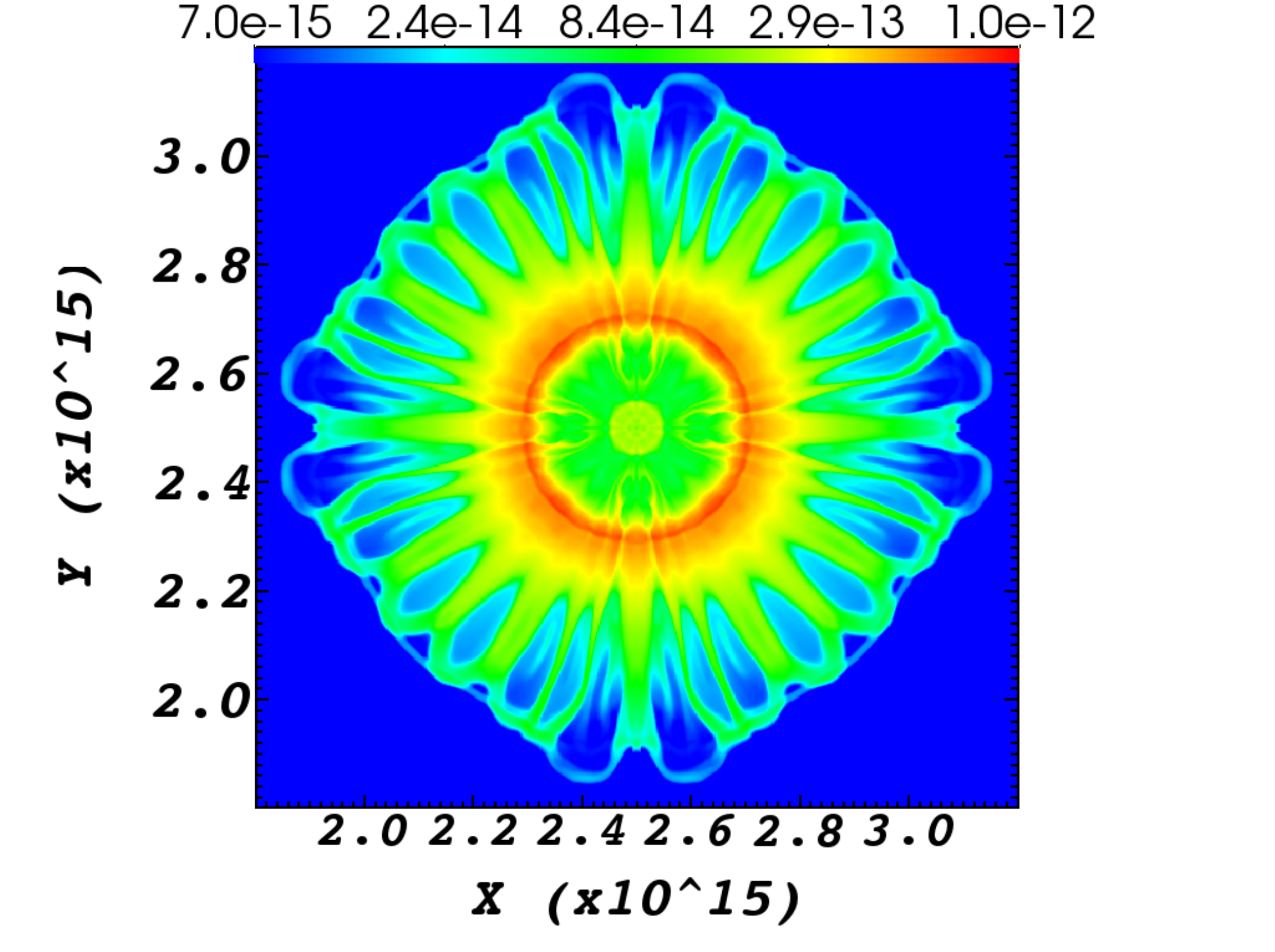}}
 \hfill
\caption{   Like Fig. \ref{fig:g4}  but for the $\gamma=4/3$ case. } \label{fig:g6}

\end{figure}

We turn to the results obtained for the $\gamma=4/3$ case, which
mimics a more efficient radiate cooling than the previous case. In
Fig. \ref{fig:g5} we present the time evolution of the density in
the meridional plane at four times, and in Fig. \ref{fig:g6} we
present the flow in two planes parallel to the equatorial plane
and at two times, like in Fig. \ref{fig:g4}. We note that there
are slight differences between these results to those obtained for
the $\gamma=5/3$ case. An interesting feature in this run is seen
in the lower panels (at $94$ and $140$ days) of Fig. \ref{fig:g5}.
This is the appearance of two arcs in the most dense region in the
equatorial plane. One arc is above and one below the equatorial
plane in the figure. Such a split of the denser region can enhance
the partition of the equatorial ring into clumps at later times,
even clumps that reside entirely only on one side of the
equatorial plane. We note the development of instabilities at the
later times in both planes presented in Fig. \ref{fig:g6}, most
prominently in panel (b). These will later form a clumpy ring.

\begin{figure}
\begin{center}
\includegraphics[height=4.2in, width=170mm]{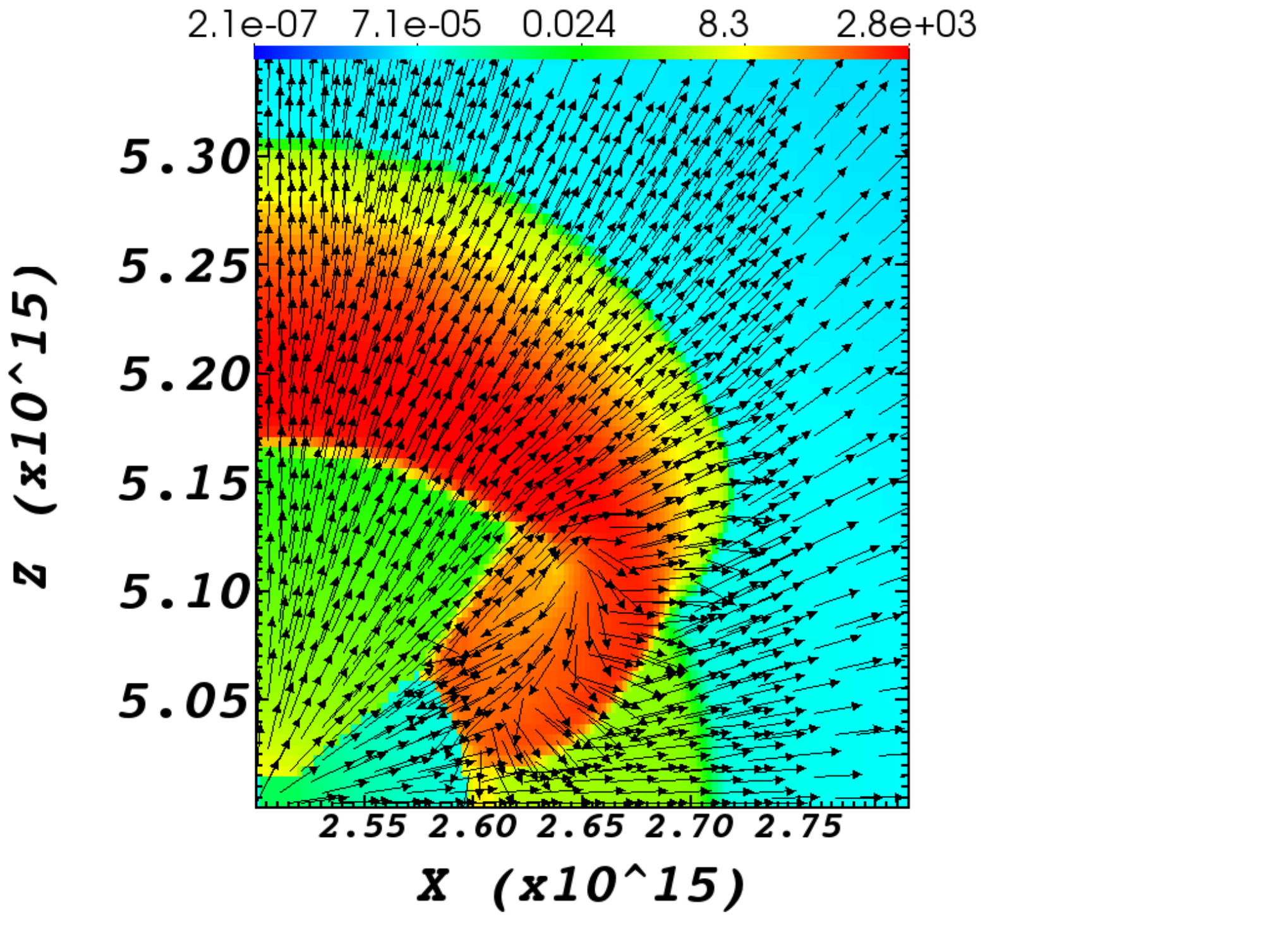}
\caption{  The pressure map in the meridional plane  at $47$ days,
for the $\gamma=5/3$ run. Arrows depict the flow direction but not
the magnitude. Density color coding is in units of $\erg
\cm^{-3}$. Units on the axes are in $\cm$. This figure emphasize
the formation of a vortex center in this plane on $(2.64,5.11)
\times10^{15} \cm$. This vortex compresses the equatorial flow. }
 \label{fig:pressure}
\end{center}
\end{figure}

To follow the compression of the equatorial flow we show in Fig.
\ref{fig:pressure} the pressure map in the meridional plane at
$47$~days for the $\gamma=5/3$ run. Arrows depict the flow
direction but not the magnitude. We clearly see a vortex in the
flow that pushes the material toward the equatorial plane. This
makes the ring to be denser and enhances the instability.

\subsection{Varying parameters}
\label{subsec:varying}

\begin{figure}
\subfigure[$t=140$~days]{\includegraphics[height=2.8in,width=3.35in,angle=0]{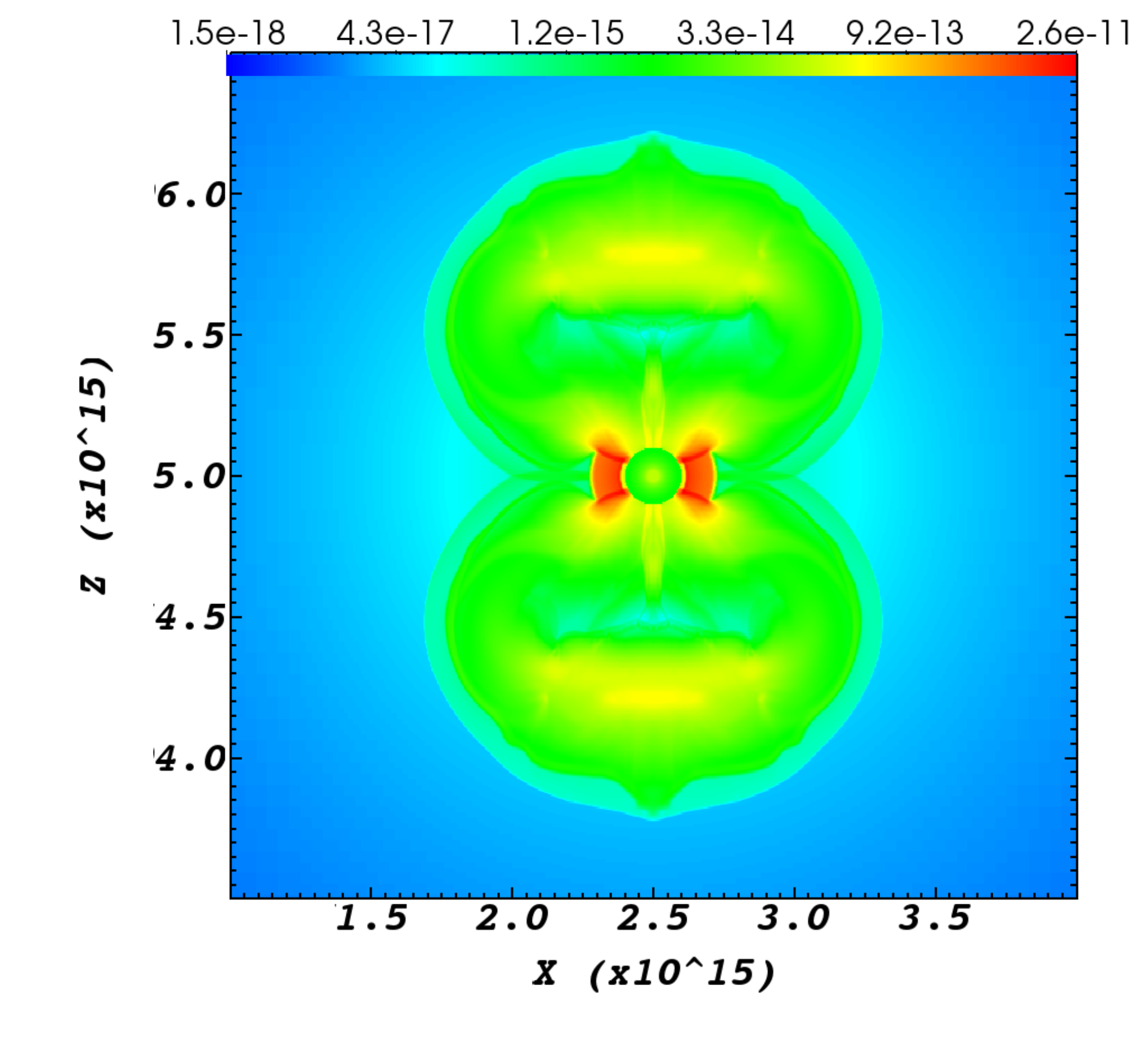}}
\hskip -0.7cm
\subfigure[$t=140$~days]{\includegraphics[height=2.8in,width=3.35in,angle=0]{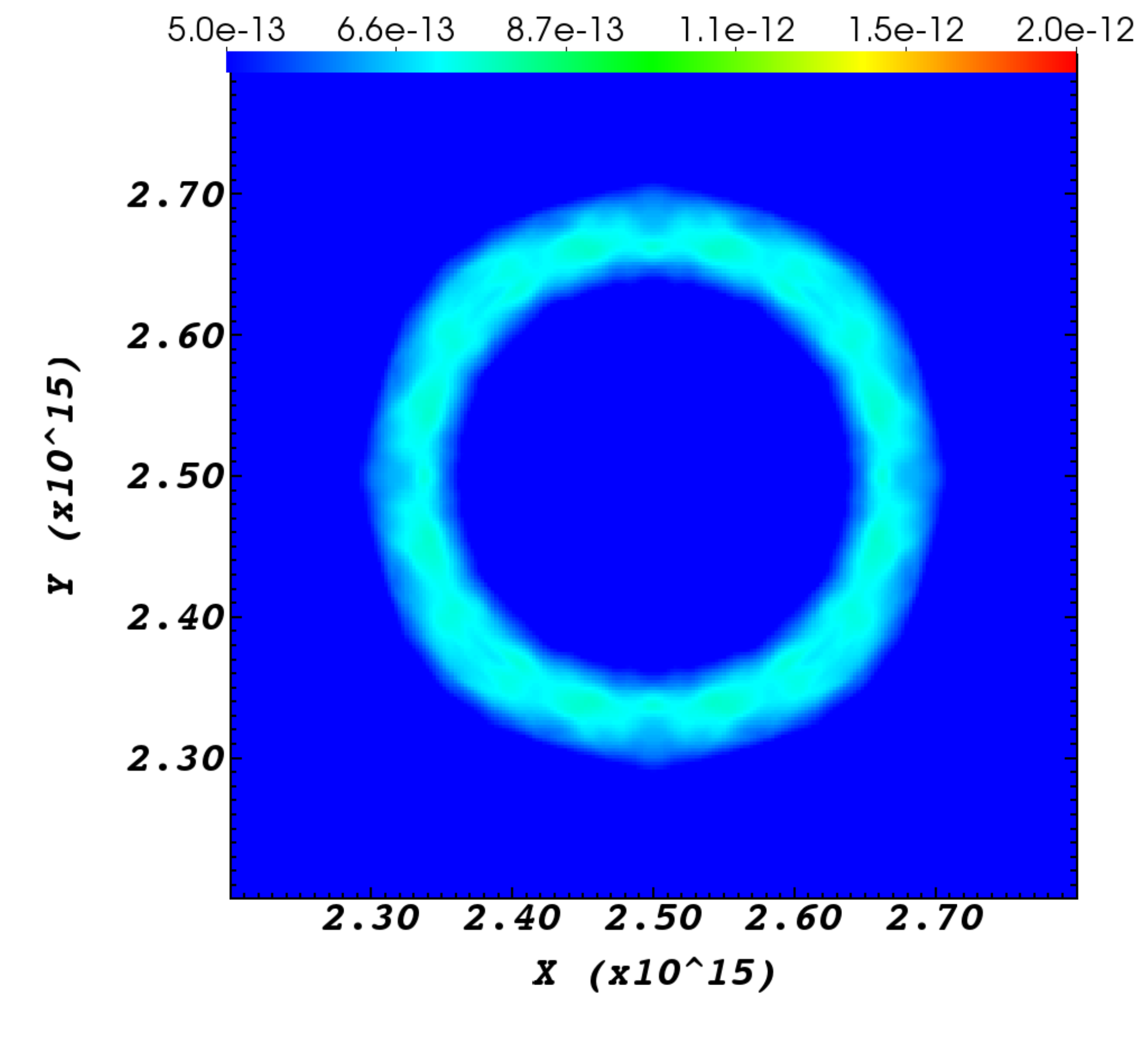}}

\caption{The density maps in the meridional plane $xz$
(left), and in a plane parallel to the equatorial plane and
$\Delta z= 0.1 \times 10^{15} \cm$ above it (right).  The half opening angle of the jets is 30 degrees.
All other parameters are as in the run presented in Figs.~\ref{fig:g3} and \ref{fig:g4}.
Color coding is in $\g \cm^{-3}$ and units on the axes are in $\cm$. }
 \label{fig:theta30}
\end{figure}

\begin{figure}
\subfigure[$t=115$~days]{\includegraphics[height=2.8in,width=3.35in,angle=0]{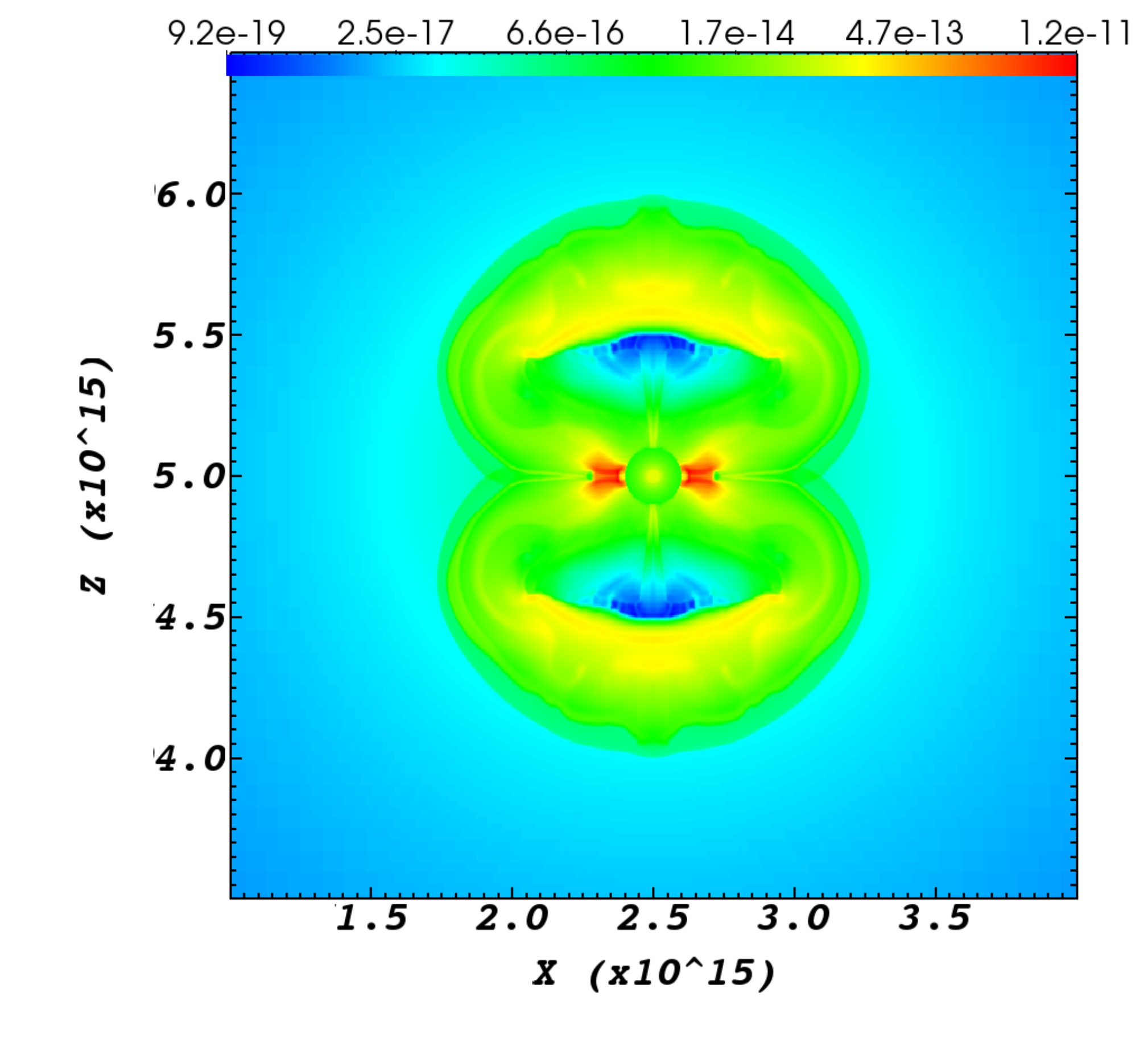}}
\hskip -0.7cm
\subfigure[$t=115$~days]{\includegraphics[height=2.8in,width=3.35in,angle=0]{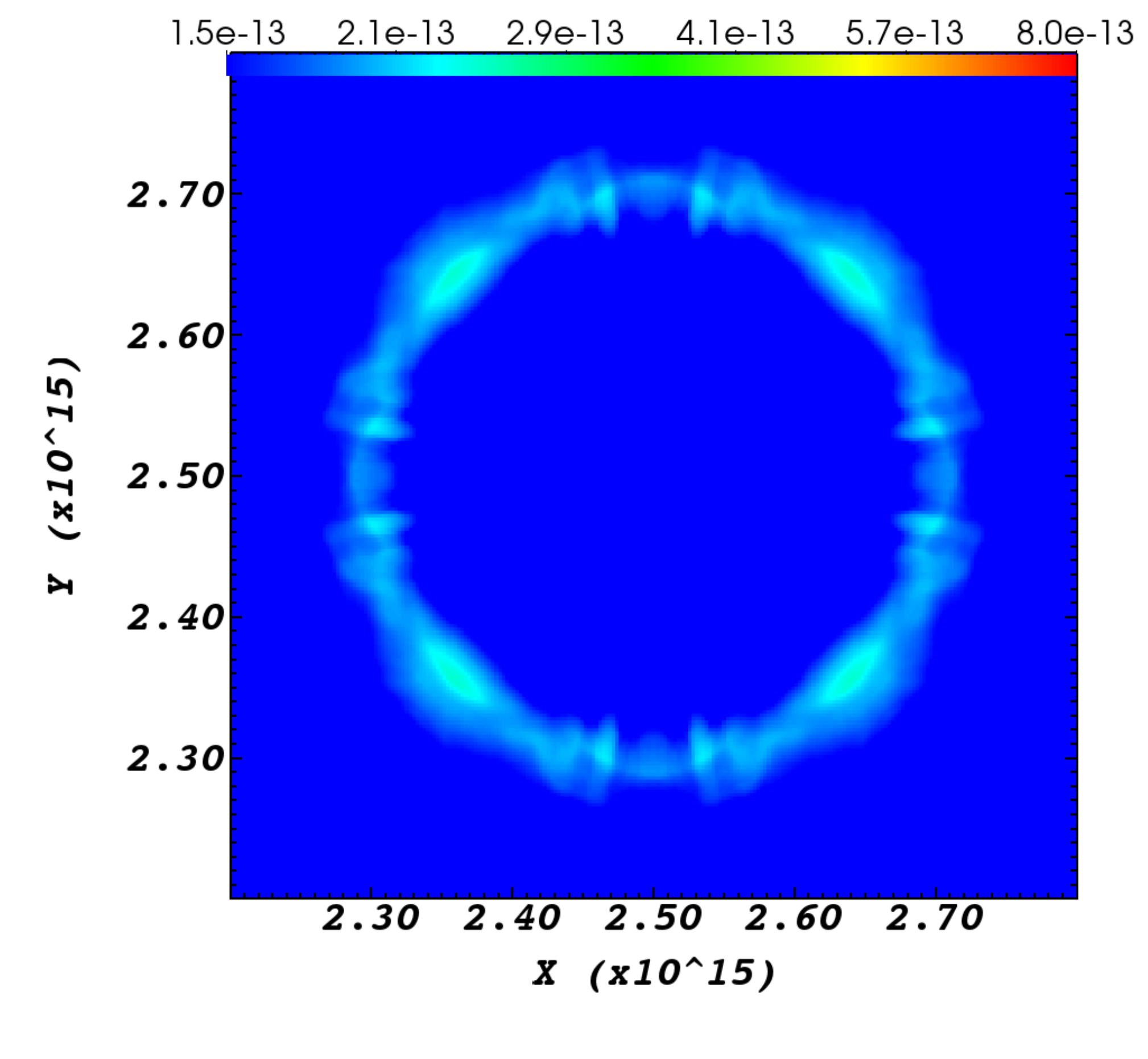}}

\caption{Like Fig. \ref{fig:theta30} but the half-opening angle of the jets is 50 degrees and the initial density of the shell is third of that used before.
 } \label{fig:theta50_third}
\end{figure}

\begin{figure}
\subfigure[$t=195$~days]{\includegraphics[height=2.8in,width=3.25in,angle=0]{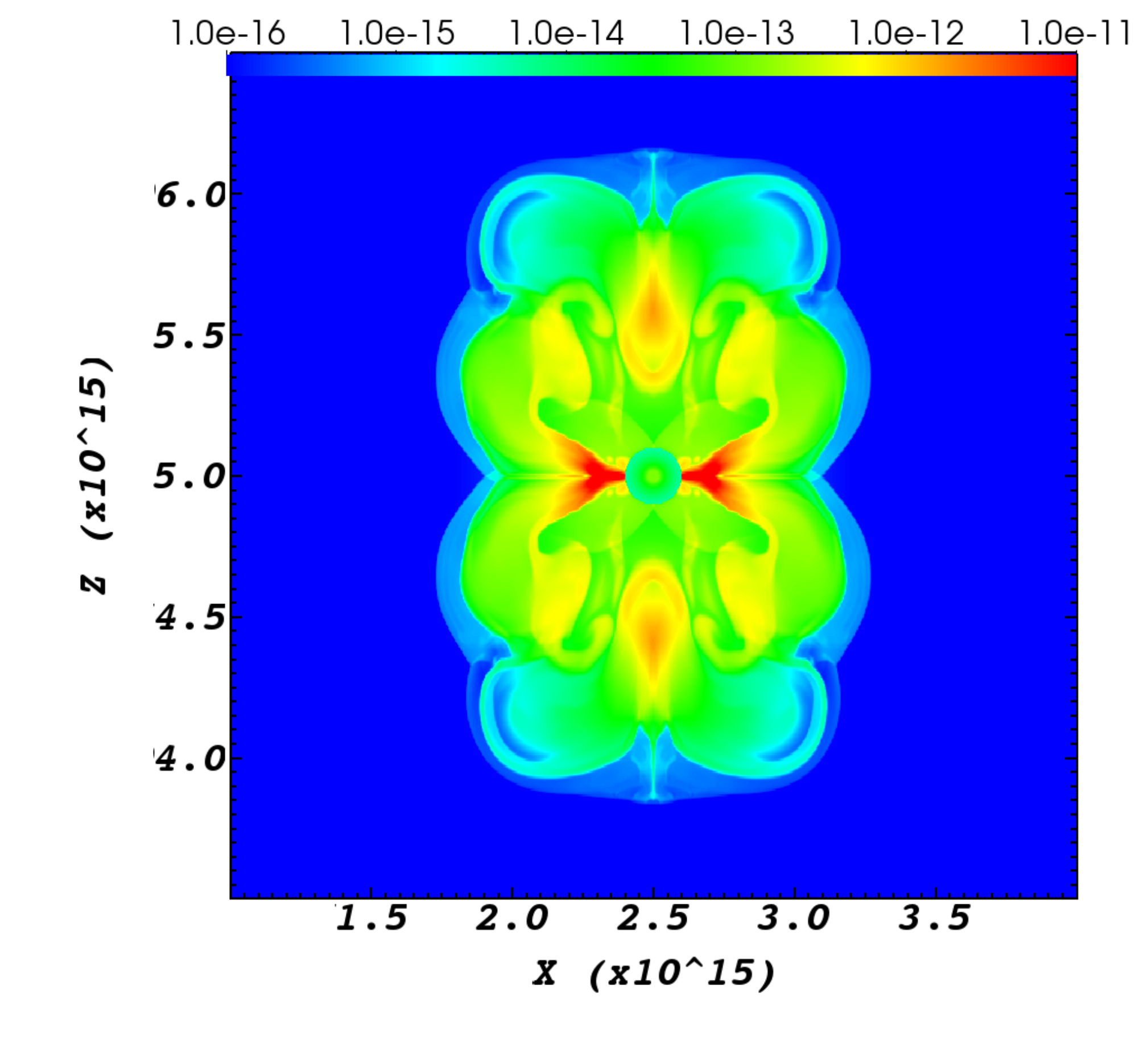}}
\hskip 0.4cm
\subfigure[$t=195$~days]{\includegraphics[height=2.8in,width=3.25in,angle=0]{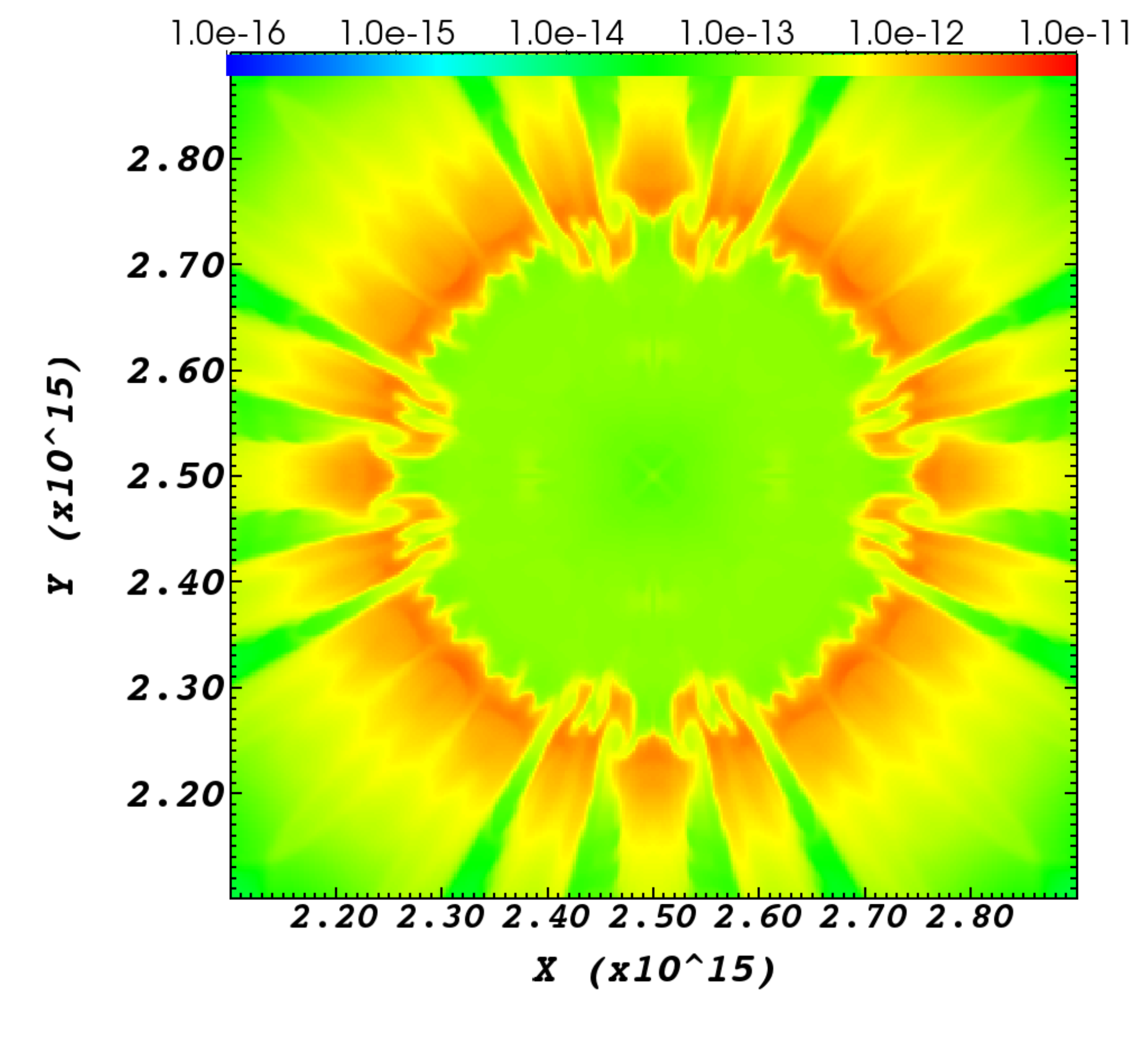}}

\caption{Like Fig. \ref{fig:theta30} but the half-opening angle of the jets is 50 degrees and the initial density of the shell is three times of that used before.  } \label{fig:theta50_three}
\end{figure}

 To examine the sensitivity of our results we performed
numerical simulations with some other initial conditions. In
Fig.~\ref{fig:theta30} we show the flow for a case where the
initial half-opening angle of the jets is 30 degrees instead of 50
degrees. All other parameters are as in the run presented in
Figs.~\ref {fig:g3} and \ref{fig:g4}. We also simulated two cases
with a half-opening angle of 50 degrees, but with an initial shell
density of third that used in previous runs, for which the results
are presented in Fig.~\ref{fig:theta50_third}, and with an initial
shell density three times as large that used in previous runs, for
which the results are presented in Fig. \ref{fig:theta50_three}.
In all cases the images clearly show the development of an
equatorial ring.    

\subsection{Summarizing the conditions for ring formation}
\label{subsec:ring}

 A key process for the presently proposed model for
the formation of a clumpy equatorial ring is that the CSM gas is
compressed toward the equatorial plane by the jets, and that it
survives for a long time. For that to occur the following
conditions must be met.
 \newline
(1) Cooling time. Since the initial momentum of each jet is away
from the equatorial plane, to exert the force toward that plane,
the jet must be shocked and form a hot bubble. For the hot bubble
to exert enough pressure its cooling time must be long. Either the
radiative emissivity is low, as in \cite{AkashiSoker2008}, or the
photon diffusion time out, $\tau_{\rm diff}$, is not much shorter
than the flow time, $t_{\rm f}$, as is the case here. The ratio
$\tau_{\rm diff}/ t_{\rm f}$ is given in equation (8) of
\cite{AkashiSoker2008}.
\newline
(2) No holes in the CSM. The compression of the CSM toward the
equatorial plane must occur before the hot bubble pressure
decreases too much. To not lose pressure by gas leakage, there
should be no `holes' in the CSM through which the hot bubble gas
can escape.
\newline
(3) No hole drilling. Another condition for the bubble pressure
not to decrease too rapidly is that the expansion along the polar
direction should not be much faster than in other directions. This
translates for the jets not to be too narrow. With a half opening
angle of $\alpha = 30^\circ$ we still get a clear ring. The exact
limit depends on the other parameters of the jets and the density
profile of the CSM. For example, in a case with a little CSM mass
along the polar directions and narrow jets, the jets will drill a
hole through the CSM, and the hot bubble gas will leak. For the
CSM and outflow parameters used here jets with $\alpha \ga 0.1 \pi
\simeq 20^\circ$ can lead to a ring formation.
\newline
(4) Sufficient energy in the shocked jets. To influence the
equatorial gas there should be enough energy in the bubble to
compress the CSM near the equatorial plane. In our simulations the
total kinetic energy of the jets were about three orders of
magnitude more than the energy (thermal and kinetic) of the CSM.
However, inspection of the velocity maps show that the fast
outflow takes place away from the equatorial plane. Only a small
fraction of the jets' energy goes to the compression. In our
simulations a clear ring develops early on, even for narrow jets
and for dense CSM. We estimate that it is sufficient that the
energy in wide-fast jets be more than an order of magnitude of
that in the CSM for a ring to form. Narrower jets require more
energy for the hot bubble they inflate to form a ring.
\newline
(5) No destruction by a late central wind. In some cases, like in
SN~1987A and in PNe, after the short interaction, the central star
can blow a fast spherical wind. Although the wind is tenuous, it
can last for thousands of years. This fast wind can clear the
lower density regions of the bipolar bubbles, and leave a large
contrast between the equatorial ring and the rest of the nebula.
As well, it can lead to the formation of two rings, one above and
one below the equatorial plane. It is required that the fast wind
does not destroy the ring. Our study of a flow with a spherical
fast wind will be published in the near future.

\section{SUMMARY }
\label{sec:summary}

We studied the formation of expanding equatorial rings around
evolved giant stars (RGB, AGB, and RSG stars). Such rings can
reveal themselves when ionized at later times by the central star,
either a central star of a PN or an exploding massive star. Based
on many similarities between rings around massive stars, that are
progenitors of core collapse supernovae (CCSNe), and rings in PNs,
we assumed that the ring formation mechanism is similar in all
these types of evolved stars (section \ref{sec:intro}). The main
similarity we referred to is the presence of a bipolar nebula in
most of these systems.

In section \ref{sec:L2} we examined the formation of the ring by
mass loss from the second Lagrangian point ($L_2$) beyond the
compact companion to the giant star. This process is most likely
to occur when the system is in a Darwin unstable phase (section
\ref{sec:Darwin}). We found that this process might indeed take
place \citep{Livioetal1979}, but that extra energy must be given
to the mass leaving from $L_2$. A second problem might be to
maintain the flow of gas from $L_2$ parallel to the equatorial
plane. The gas is close to the binary system, hence hot. It is
expected to expand away from the equatorial plane, making the
formation of a dense equatorial ring questionable. A third problem
might be the formation of a clumpy ring, as the outflow from $L_2$
occurs over many orbital periods and the ring is expected to be
smooth.

In section \ref{sec:hydro} we studied the compression of an
equatorial ring by bipolar jets, as suggested by
\cite{SokerRappaport2000}. We found that the process leads to the
formation of a dense and clumpy equatorial ring. We therefore
suggest that clumpy equatorial rings, such as in the Necklace PN
and in AN~1987A, as well as many other similar objects, are formed
by this process. In most cases the jets are launched from an
accretion disk around a compact companion.

In this first study we did not follow the ring evolution beyond
the short time of its formation. In most, or even in all, of the
systems with clumpy rings mentioned in section \ref{sec:intro}, a
fast tenuous wind is blown after the intensive mass loss episode.
This fast wind might clean the low-density regions, leaving behind
only the dense regions, such as the equatorial ring. This might
even form polar rings, such as the outer rings in SN~1987A. The
later evolution of the ring after the fast tenuous wind has turned
on is the subject of a forthcoming paper.

\bigskip
{\bf ACKNOWLEDGEMENT}
\newline
 This research was supported by the Asher Fund for
Space Research at the Technion, and the US-Israel Binational
Science Foundation.

\end{document}